\documentclass[12pt]{iopart}
\usepackage{color}
\usepackage{graphicx}
\usepackage{amssymb}

\eqnobysec

\usepackage{citesort}

\begin{document}
\title[Phase control of entanglement and quantum steering]{Phase control of entanglement and quantum steering in a three-mode optomechanical system}
\author{F. X. Sun$^{1}$, D. Mao$^{1}$, Y. T. Dai$^{1}$, Z. Ficek$^{2,3}$, Q. Y. He$^{1,4,*}$ and Q. H. Gong$^{1,4}$}
\address{$^{1}$State Key Laboratory of Mesoscopic Physics, School of Physics, Peking University, Collaborative Innovation Center of Quantum Matter, Beijing 100871, China \\ $^{2}$The National Centre for Applied Physics, KACST, P.O. Box 6086, Riyadh 11442, Saudi Arabia \\ 
$^{3}$ Institute of Physics, University of Zielona G\'{o}ra,  65-516 Zielona G\'{o}ra, Poland \\$^{4}$Collaborative Innovation Center of Extreme Optics, Shanxi University, Taiyuan 030006, China \\ $^*$ E-mail: qiongyihe@pku.edu.cn}

\begin{abstract}
The theory of phase control of coherence, entanglement and quantum steering is developed for an optomechanical system composed of a single mode cavity containing a partially transmitting dielectric membrane and driven by short laser pulses. The membrane divides the cavity into two mutually coupled optomechanical cavities resulting in an effective three-mode closed loop system, two field modes of the two cavities and a mechanical mode representing the oscillating membrane. The closed loop in the coupling creates interfering channels which depend on the relative phase of the coupling strengths of the field modes to the mechanical mode. Populations and correlations of the output modes are calculated analytically and show several interesting phase dependent effects such as reversible population transfer from one field mode to the other, creation of collective modes, and induced coherence without induced emission. We find that these effects result from perfect mutual coherence between the field modes which is preserved even if one of the modes is not populated. The inseparability criterion for the output modes is also investigated and we find that entanglement may occur only between the field modes and the mechanical mode. We show that depending on the phase, the field modes can act on the mechanical mode collectively or individually resulting, respectively, in tripartite or bipartite entanglement. In addition, we examine the phase sensitivity of quantum steering of the mechanical mode by the field modes. Deterministic phase transfer of the steering from bipartite to collective is predicted and optimum steering corresponding to perfect EPR state can be achieved. These different types of quantum steering can be distinguished  experimentally by measuring the coincidence rate between two detectors adjusted to collect photons of the output cavity modes. In particular, we find that the minima of the interference pattern of the coincidence rate signal the bipartite steering, while the maxima signal the collective steering.

\end{abstract}

\submitto{\NJP}
\maketitle

\section{Introduction}

It is well know that optical coherence and quantum entanglement between two systems can occur when the source systems are prepared in a superposition state~\cite{mandel1995optical,ficek2005quantum}. Especially, the mutual coherence is described by the first-order correlation between two undistinguishable pathways or channels and can determine the phase of a radiating unknown system relative to the phase of the other known system.  On the other hand, quantum entanglement determines inseparability of quantum systems, and has been recognized as one of the most intrinsic features of quantum mechanics with many useful applications ranging from quantum cryptography, quantum metrology, to quantum computation~\cite{RevModPhys.81.865,RevModPhys.81.1727,RevModPhys.86.419}. A particular interest is devoted to a special type of entanglement called quantum steering~\cite{wiseman2007steering,jones2007entanglement,Reportsteering}. The concept of quantum steering was originally introduced by Schr\"{o}dinger~\cite{schrodinger1935discussion} to explore the fact that entanglement would allow one to remotely steer or pilot the state of a distant system, as considered in the original Einstein-Podolsky-Rosen (EPR) paradox~\cite{einstein1935can}. The EPR steering allows two parties to verify the shared entanglement even if one measurement device is untrusted, which makes it an essential resource for one-sided device independent quantum cryptography~\cite{branciard2012one,walk2016experimental,gehring2015implementation,gallego2015resource,NP2015,Guo16}, one-way quantum computing~\cite{Pan15,gheorghiu2017rigidity}, secure quantum teleportation~\cite{he2015classifying,LCMnpj}, and subchannel discrimination~\cite{piani2015necessary}. 

Recent studies have revealed that coherence is closely related to entanglement and quantum steering. It was pointed out by Suzuki {\it et. al}~\cite{suzuki2010entanglement} that entanglement can be detected from interference fringes in atom-photon systems. It has also been shown that the coherence in a system and entanglement between that system and another initially incoherent one are quantitatively, or operationally, equivalent~\cite{streltsov2015measuring}. The power of quantum steering for the generation of coherence has also been demonstrated~\cite{hu2016quantum,PhysRevA.95.010301}. It has been shown that the presence of mutual coherence among two systems may have a distractive effect on the creation of entanglement between these systems~\cite{sun2012first}. On the other hand, it has been shown that in a tripartite system the mutual coherence between two parties may help to collectively steer the third party~\cite{wang2015efficient}.  

The hope to demonstrate entanglement and quantum steering in macroscopic systems has encouraged research on optomechanical systems. The successful achievement of cooling of nanomechanical oscillators to near their ground states~\cite{o2010quantum,chan2011laser,teufel2011sideband,teufel2011circuit} makes possible to use optomechanical systems to study quantum mechanical effects in mesoscopic massive systems~{\cite{safavi2012observation,li2016quantum,kippenberg2008cavity,aspelmeyer2010quantum,carmele2014opto}. It has been shown that optomechanical systems can be used to generate entangled states between a mechanical oscillator and an optical (microwave) field~\cite{aspelmeyer2014cavity,ferreira2006macroscopic,paternostro2007creating,vitali2007entangling,bhattacharya2008entanglement,hartmann2008steady,wallquist2010single,barzanjeh2011entangling,zhou2011entanglement,hofer2011quantum,he2013einstein,kiesewetter2014scalable,yin2009generating,vanner2011pulsed}. For example, an elegant electromechanical experiment has reported the observation of a bipartite entanglement of a microwave field with a mechanical oscillator~\cite{palomaki2013entangling}. Further theoretical studies have considered the generation of  entanglement and quantum steering in tripartite optomechanical systems, where two independent modes can be made entangled when one of the mode is coupled to the intermediate mode by a parametric interaction and the other is coupled by a linear-mixing interaction~\cite{he2014einstein,genes2008simultaneous,genes2008emergence,sun2012first,wang2013reservoir,xiang2015detection,wang2014role}. 

	Apart from the extensive efforts toward the creation of entangled states, it is crucial for quantum information processing to be able to control the creation and evolution of the entangled states. Recent research on phase dependent systems has addressed the problem of controlling the entanglement in two qubits systems~\cite{malinovsky2004quantum,malinovsky2006phase}, a triple spin qubit~\cite{nakajima2016phase}, and quantum parametric oscillators~\cite{gonzalez2015generation}. Other simple systems have demonstrated the optical nonreciprocal behaviour induced by the phase difference between the coupling constants in a fully coupled tripartite optomechanical system~\cite{xu2015optical}, or in a multimode on-chip electromechanical system~\cite{barzanjeh2017mechanical}, and novel quantum interference effects and correlations arising from the interplay and competition between different excitation channels~\cite{PhysRevA.95.023832}.

In this paper we explore the possibility of the phase controlled generation and transfer of coherence, entanglement and quantum steering in a closed loop multimode system. We propose to consider a close loop coupling between modes in optomechanics. In particular, we examine phase dependent dynamics an optomechanical system composed of a single mode cavity containing a dielectric membrane in its interior. We assume that the membrane can behave as a partly transmitting and reflecting mirror which divides the cavity into two mutually coupled optomechanical cavities. We examine phase properties of the coherence functions, the first-order coherence and anomalous correlation functions involved in the generation of entanglement between modes. We then propose to use such phase dependent functions in the controlled generation and transfer of entanglement among the modes. Specifically, we show that the presence of the mutual coherence between the field modes may result in perfect entanglement between the mechanical and one of the field modes. By varying the relative phase of the coupling constants of the field modes to the mechanical mode, one can achieve the perfectly entangled tripartite state involving the mechanical mode and a linear superposition of the field modes. In addition, we demonstrate the crucial role of the phase in the quantum steering of the mechanical mode by the field modes, the achievement of perfectly steerable EPR state and the phase controlled distinguishability of the collective and bipartite steering.  

The paper is organized as follows. In section~\ref{sec.2} we describe in detail our model and discuss the method of the normalized temporal pulse-shape amplitudes used to solve the appropriate equations of motion describing the evolution of the mechanical mode and the cavity field operators. These equations are then used to find the analytical expressions for the population of the modes and for the first-order correlations and anomalous correlation functions. We apply the expressions in section~\ref{sec.3} to determine the mutual coherence between the modes and the phase dependent transfer of the population between the modes. We also present there results concerning the degree of coherence, visibility of the interference fringes, distinguishability of the modes and the phase dependence of the coincidence rate between two detectors adjusted to collect photons of the output modes of the two cavities. 
Section~\ref{sec.4} is devoted to the discussion of the inseparability criterion. We present the analytical form of the separability parameter and discuss in details its phase properties. In section~\ref{sec.5} we focus on steering properties of the modes, discuss the possibility of achieving a perfectly steerable EPR state between the modes, and suggest a measurement technique to distinguish between the bipartite and collective steerings. We summarize our results in section~\ref{sec.6}. Finally, in the Appendix, we give results for the separability parameters evaluated for excitation of the cavity modes with a noisy laser.

\section{The model}\label{sec.2}

We consider a system composed of a single mode cavity containing a dielectric membrane in its interior, as illustrated in figure~\ref{fig:scheme}(a). The membrane divides the cavity into two cavities resulting in a system effectively behaving as a three-mode optomechanical system: two field modes of the two cavities plus a mechanical mode representing the vibrating membrane. The field modes and the mechanical mode are treated as quantized: $a_{j}\, (a_{j}^{\dag}),\, j=1,2$, is the annihilation (creation) operator for the mode of cavity $j$, and $c\, (c^{\dag})$ is the annihilation (creation) operator for the mechanical mode. The modes of these two cavities have frequencies $\omega_1$ and $\omega_2$ depending on the position of the membrane relative to the position of the cavity mirrors. The mechanical mode has a frequency $\omega_m$. The field modes are driven by external detuned short laser pulses of equal duration time $\tau$, equal frequencies $\omega_{L}$, and electric field amplitudes $E_{1}(t)=E_{01}\exp[-i(\omega_{L}t +\varphi_{01})]$ and $E_{2}(t)=E_{02}\exp[-i(\omega_{L}t +\varphi_{02})]$. The laser pulses are derived from the same laser to ensue a constant phase difference $\delta\varphi_0 =\varphi_{01}-\varphi_{02}$.  The thickness of the membrane is very small that it can be treated as a nonabsorptive partly reflecting and transmitting mirror. The possibility of the field transmission through the membrane results in a direct linear coupling of the modes $a_{1}$ and $a_{2}$ with a strength $J$. This coupling together with the coupling to the mechanical mode creates a three-mode closed linkage or ``loop'' coupling of the modes of the system, as illustrated in~figure~\ref{fig:scheme}(b). The presence of this closed loop will give rise to interference effects and phase dependence of the dynamics of the modes.
\begin{figure}[h]
\centering
\includegraphics[width=0.8\textwidth]{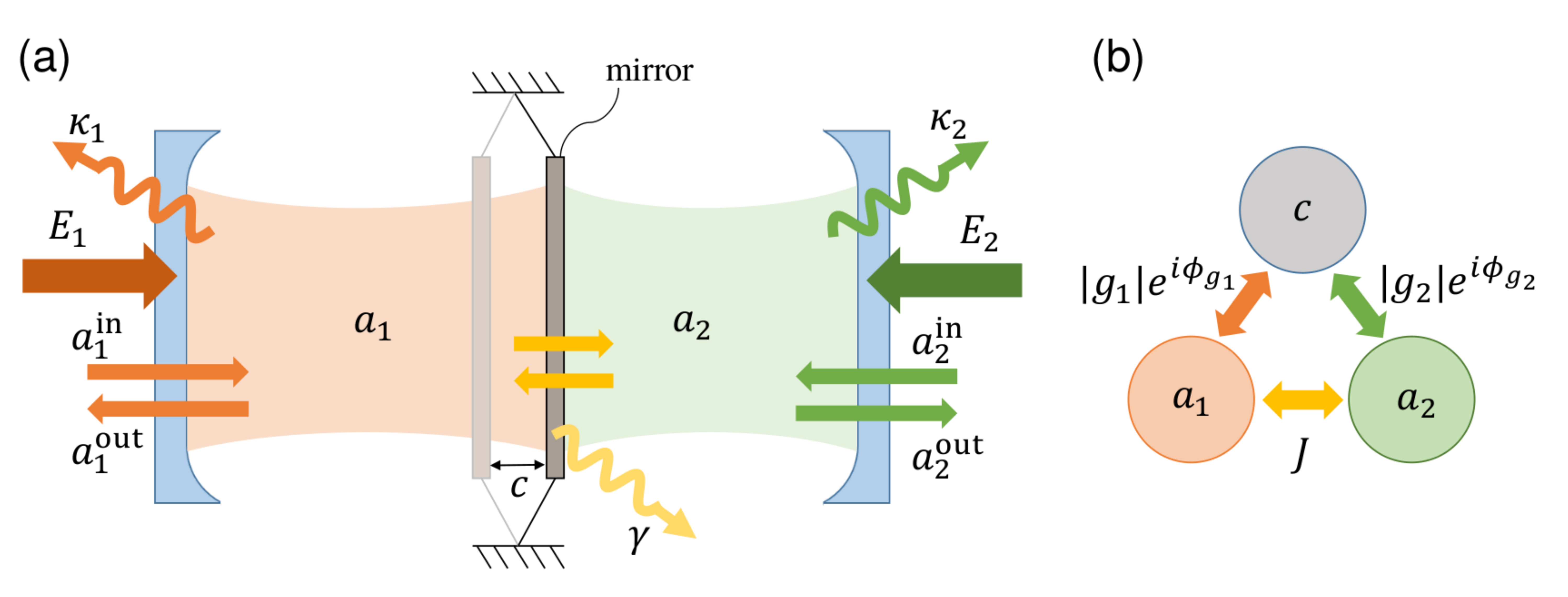}
\caption{(a) Schematic diagram of a tripartite optomechanical system. The two field modes and moveable mirror are represented by annihilation operators $a_{1,2}$ and $c$. Here, $a_{1,2}^{\rm in}$ and $a_{1,2}^{\rm out}$ denote input and output cavity fields. (b) The different coupling strength and phases are introduced between two modes of three.}
\label{fig:scheme}
\end{figure}

The Hamiltonian of this system may be written as 
\begin{eqnarray}
H &=& \hbar\omega_{1}a^{\dag}_{1}a_{1}+\hbar\omega_{2}a^{\dag}_{2}a_{2}+\hbar\omega_{m}c^{\dag}c+\hbar J(a_{1}^{\dag}a_{2}+ a_{2}^{\dag}a_{1}) \nonumber \\
&&+\hbar (g_{0,1}a^{\dag}_{1}a_{1} + g_{0,2}a^{\dag}_{2}a_{2})(c^{\dag}+c) \nonumber\\
&&+i\hbar[E_{1}(t)a_{1}^{\dag} +E_{2}(t)a_{2}^{\dag} - {\rm H.c.}] ,\label{H}
\end{eqnarray}
where $g_{0,j}$ is the single-photon coupling of the mode $j$ to the mechanical mode. The Hamiltonian (\ref{H}) may be met in typical experimental optomechanical systems with a membrane in a Fabry-Perot cavity~\cite{aspelmeyer2014cavity}. 

Using the Heisenberg equation of motion and Hamiltonian (\ref{H}), we obtain the following equations of motion for the annihilation operators
\begin{eqnarray} 
	\dot{a}_{1}^{r} &= & -(\kappa_{1}+i\Delta_{1})a_{1}^{r}-ig_{0,1}a_{1}^{r}(c+c^{\dagger}) -iJa_{2}^{r}e^{i\delta\varphi_0}+E_{01}-\sqrt{2\kappa_{1}}a_{1}^{{\rm in,r}} , \nonumber\\
	\dot{a}_{2}^{r} &=& -(\kappa_{2}+i\Delta_{2})a_{2}^{r}-ig_{0,2}a_{2}^{r}(c+c^{\dagger})-iJa_{1}^{r}e^{-i\delta\varphi_0}+E_{02}-\sqrt{2\kappa_{2}}a_{2}^{{\rm in,r}} , \nonumber\\
	\dot{c} &=& -(\gamma+i\omega_{m})c-ig_{0,1}a_{1}^{r\dagger}a_{1}^{r}-ig_{0,2}a_{2}^{r\dagger}a_{2}^{r}-\sqrt{2\gamma}c^{in} ,\label{II2}
\end{eqnarray} 
along with the corresponding equations for the creation operators. Here, $a_{j}^{r} = a_{j}\exp\{i\left[ \omega_{L}t+\varphi_{0j}\right]\}$, $a_{j}^{{\rm in,r}}= a_{j}^{{\rm in}}\exp\{i\left[ \omega_{L}t+\varphi_{0j}\right]\}$, and $\Delta_{j}=\omega_{j}-\omega_{L}$ is the detuning of the laser frequency from the frequency of the mode $j$. In writing equation~(\ref{II2}) we have included relaxation terms of the modes: $\kappa_{j}$ is the damping rate of the field mode $j$, and $\gamma$ is the damping rate of the mechanical mode. We have also included the noises of the input modes $a^{\rm in}_{j}$ and $c^{\rm in}$ arising from the coupling of the modes to their surrounding environments. Note that the dependence of the equations of motion (\ref{II2}) on the phase difference $\delta\varphi_{0}$ only arises from the linear coupling of the two cavity modes. For the statistics of the input modes we assume that the input modes of the cavities are in the ordinary vacuum state characterized by the correlation function $\langle a^{\rm in}_{j}(t)(a_{j}^{\rm in})^{\dagger}(t')\rangle =  \delta(t-t')$, and the input mode to the vibrating membrane is in a thermal vacuum states characterized by the correlation functions $\langle c^{\rm in}(t)(c^{\rm in})^{\dagger}(t')\rangle = [n(\omega_{m})+1]\delta(t-t')$ and $\langle (c^{\rm in})^{\dag}(t)c^{\rm in}(t')\rangle = n(\omega_{m})\delta(t-t')$, where $n(\omega_{m})=1/[\exp(\hbar\omega_{m}/k_{B}T)-1]$ is the average number of thermal phonons at the frequency of the mechanical mode, $k_B$ is the Boltzmann constant and $T$ is the temperature of the environment.

Assuming $E_{0j} \gg \kappa_{j},\gamma$, the equations of motion (\ref{II2}) may be solved by the linearization approach~\cite{braunstein2012quantum}. In this approach, we write the operators of the system as composed of their steady-state mean values and a small fluctuation around the steady-state. The steady-state mean values of the operators, $\langle a_{j}\rangle \equiv \alpha_{j}$ and $\langle c\rangle\equiv \chi$, are obtained by setting the left-hand sides of equation~(\ref{II2}) to zero yielding  
\begin{eqnarray}
\alpha_{1} &=& \frac{(\kappa_{2}+i\Delta'_{2})E_{01}e^{i\varphi_{01}} -iJE_{02}e^{i\varphi_{02}}}{(\kappa_{1}+i\Delta'_{1})(\kappa_{2}+i\Delta'_{2})+J^{2}}, \ \ \alpha_{2} = \frac{(\kappa_{1}+i\Delta'_{1})E_{02}e^{i\varphi_{02}} -iJE_{01}e^{i\varphi_{01}}}{(\kappa_{1}+i\Delta'_{1})(\kappa_{2}+i\Delta'_{2})+J^{2}} , \nonumber\\
\chi &=& \frac{-i(g_{0,1}|\alpha_1|^2 +g_{0,2}|\alpha_2|^2)}{\gamma+i\omega_m} ,\label{II3}
\end{eqnarray}
where $\Delta'_{j} = \omega_{j}^{\prime} -\omega_{L}$, with $\omega_{j}^{\prime} = \omega_{j} + g_{0,j}(\chi+\chi^{\ast})$.

Using the expansions $a_j \rightarrow \alpha_j + \delta a_j$ and $c \rightarrow \chi+ \delta c$, we obtain the equations of motion for the fluctuation parts of the operators. Keeping the linear terms only, the equations of motion are of the form
\begin{eqnarray}
	\delta \dot{a}_{1}^{r} &=& -(\kappa_{1}+i\Delta'_{1})\delta a_{1}^{r}-ig_1e^{i\varphi_{01}}(\delta c+\delta c^\dagger)-iJe^{i\delta\varphi_0}\delta a_{2}^{r}-\sqrt{2\kappa_{1}}a^{\rm in,r}_{1}, \nonumber\\
\delta \dot{a}_{2}^{r} &=& -(\kappa_{2}+i\Delta'_{2})\delta a_{2}^{r}-ig_2e^{i\varphi_{02}}(\delta c+\delta c^\dagger)-iJe^{-i\delta\varphi_0}\delta a_{1}^{r}-\sqrt{2\kappa_{2}}a^{\rm in,r}_{2}, \nonumber\\
\delta \dot{c} &=& -(\gamma+i\omega_m )\delta c -i(g_1^{*}e^{-i\varphi_{01}}\delta a_1^{r}+g_1e^{i\varphi_{01}}\delta a_1^{r\dag}) \nonumber\\
&&-i(g_2^*e^{-i\varphi_{02}}\delta a_2^{r}+g_2e^{i\varphi_{02}}\delta a^{r\dag})-\sqrt{2\gamma}c^{\rm in} ,
\label{II4}
\end{eqnarray}
where $g_j=g_{0,j}\alpha_j$ is the effective optomechanical coupling strength of the mode $j$ to the mechanical mode.

It is seen from equation~(\ref{II4}) that the fluctuation operators oscillate at frequencies $\pm\omega_{m}$. Therefore, it is convenient to introduce slowly varying fluctuation operators $\delta c^m = \delta c e^{i\omega_mt},\ \delta c^{\rm in,m} = \delta c^{\rm in}e^{i\omega_mt}, \ \delta a^{r,m}_{j} = \delta a_{j}^{r}e^{-i(\omega_mt +\varphi_{0j})}$, $\delta a^{\rm in,r,m}_{j} = \delta a^{\rm in,r}_{j}e^{-i(\omega_mt +\varphi_{0j})}$, and substitute them into equation~(\ref{II4}). Discarding all terms oscillating at $2\omega_m$ based on the rotating-wave approximation, we then obtain 
\begin{eqnarray}
\dot{a}_{1} &=& -\left(\kappa_{1}+i\Delta\right)a_{1}-ig_1c^{\dag}-iJa_{2}-\sqrt{2\kappa_{1}}a^{\rm in}_{1} ,\nonumber\\
\dot{a}_{2} &=& -\left(\kappa_{2}-i\Delta\right)a_{2}-ig_2c^{\dag}-iJa_{1}-\sqrt{2\kappa_{2}}a^{\rm in}_{2} ,\nonumber\\
\dot{c} &=& -\gamma c-ig_1a_{1}^{\dag}-ig_2a_2^\dag-\sqrt{2\gamma}c^{\rm in} .
\label{II6}
\end{eqnarray}
Here we have assumed that the laser pulses are tuned to the blue sideband of the average frequency of the field modes, i.e., $\omega_L = \omega_0 + \omega_m$, where $\omega_0 = (\omega^{\prime}_1+\omega^{\prime}_2)/2$ and $\Delta = (\omega^{\prime}_1-\omega^{\prime}_2)/2$.
For clarity of the notation, we have omitted in equation~(\ref{II6}) the symbol $\delta$ and the superscripts $r$ and $m$ on the displacement operators.

From equation~(\ref{II6}) we can see that the modes $a_{1}$ and $a_{2}$ are directly coupled to each other with the strength $J$, and are also indirectly coupled to each other through the coupling to the mechanical mode with strengths $g_{1}$ and $g_{2}$, respectively. This coupling configuration forms a closed loop, as illustrated in figure~\ref{fig:scheme}(b), and therefore the dynamics of the system can exhibit phase dependent effects.  

Let us first consider the result of the coupling $J$ on the dynamics of the field modes. 
By introducing column vectors $Y_{a}=(a_{1}, a_{2})^{T}$, $Y_{a}^{{\rm in}} = (a_{1}^{\rm in}, a_{2}^{\rm in})^{T}$, and $Y_{c} = (c^{\dagger}, c^{\dagger})^{T}$, we can put the equations of motion for $a_{1}$ and $a_{2}$ into a matrix form
\begin{equation}
\dot{Y}_{a} = -\kappa Y_{a} -i\bar{M}Y_{a}-i\bar{N}Y_{c} - \sqrt{2\kappa}Y_{a}^{{\rm in}} ,\label{II7}
\end{equation}
where the matrix $\bar{M}$ describes the effects of the linear coupling $J$ between the modes and the detuning $\Delta$, whereas the matrix $\bar{N}$ governs the influence of the nonlinear coupling to the mechanical mode,
\begin{equation}
\bar{M} = \left(
\begin{array}{cc}
\Delta & J \\
J & -\Delta
\end{array}
\right) ,\quad 
\bar{N} = \left(
\begin{array}{cc}
g_{1} & 0 \\
0 & g_{2}
\end{array}
\right) .\label{II8}
\end{equation}

Diagonalization of the matrix $\bar{M}$ results in orthogonal superposition modes represented by the superposition operators 
\begin{equation}
a_{w} = a_{1}\cos\theta +a_{2}\sin\theta ,\quad 
a_{u} = a_{1}\sin\theta  -a_{2}\cos\theta ,\label{II9}
\end{equation} 
where $\cos^2\theta =1/2+\Delta/(2w)$ with $w=\sqrt{J^{2}+\Delta^{2}}$. The angle $\theta$ belongs to the interval $[0,\pi/2]$. 

Using equations~(\ref{II6}) and (\ref{II9}) one can easily find that the annihilation operators of the superposition modes satisfy the following equations of motion
\begin{eqnarray}
\dot{a}_{w} &=&-(\kappa +iw)a_{w}-ig_{w}c^{\dag}-\sqrt{2\kappa}a^{\rm in}_{w} ,\nonumber\\
\dot{a}_{u} &=&-(\kappa -iw)a_{u}-ig_{u}c^{\dag}-\sqrt{2\kappa}a^{\rm in}_{u} ,\nonumber\\
\dot{c} &=&-\gamma c -i\left(g_{w}a^\dagger_{w}+g_{u}a^\dagger_{u}\right) -\sqrt{2\gamma}c^{\rm in} ,\label{II10}
\end{eqnarray}
where $g_{w} = g_{1}\cos\theta + g_{2}\sin\theta, \ g_{u} = g_{1}\sin\theta - g_{2}\cos\theta $ are effective coupling strengths of the superposition modes to the mirror mode. Note that the superposition modes $a_{w}$ and $a_{u}$ only couple to the mechanical mode and not to each other. This coupling configuration contains no closed loops, it resembles a chain coupling sequence. However, the dynamics of this system can exhibit phase dependence as it involves phase-dependent coupling constants $g_{w}$ and $g_{u}$. 
Furthermore, the dynamics reduce to those involving two modes only if the conditions $\Delta=0$ and $g_{1}= g_{2}\equiv g$ are fulfilled. In this case, $g_{w}=\sqrt{2}g$ and $g_{u}=0$ that the mode $u$ completely decouples from the remaining modes. 

Equation (\ref{II10}) is the basic equation for calculating the dynamics of the modes. In the following we always concentrate on transient effects resulting from the excitation of the modes by short laser pulses. 

The solution of equation~(\ref{II10}) is in general complicated. A simple analytical solution arises, however, in two cases, the bad cavity limit, $\kappa \gg |g_{w,u}|$, or at a large difference between the frequencies of the superposition modes, $w\gg |g_{w,u}|$. Under such conditions, we can approximate $a_{w}$ and $a_{u}$ as slowly varying in time, and put $\dot{a}_{w}\approx 0$ and $\dot{a}_{u}\approx 0$, which leads to
\begin{equation}
a_{w} = \frac{-e^{-i\phi}}{\sqrt{\kappa^2+w^2}}(ig_w c^\dagger+\sqrt{2\kappa}a^{\rm in}_w), \ a_{u} = \frac{-e^{i\phi}}{\sqrt{\kappa^2+w^2}}(ig_u c^\dagger+\sqrt{2\kappa}a^{\rm in}_u) ,\label{II12}
\end{equation}
where $\phi=\arctan \left(w/\kappa\right)$. The equation of motion for the mechanical mode becomes
\begin{equation}
\dot{c} = (G+i\delta)c+i\sqrt{2G_w}e^{i\phi}a_w^{\rm in\dagger}+i\sqrt{2G_u}e^{-i\phi}a_u^{\rm in\dagger}-\sqrt{2\gamma}c^{\rm in} ,\label{II13}
\end{equation}
where $\delta = (|g_w|^2 -|g_u|^2)w/(\kappa^2+w^2)$, $G_{w} = |g_{w}|^{2}\kappa/(\kappa^{2} +w^{2})$, $G_{u} = |g_{u}|^{2}\kappa/(\kappa^{2} +w^{2})$, and $G = G_w+G_u-\gamma$. 
Note that the following results are all obtained in this approximation.

Before proceeding further, we briefly comment about the possible setups and the parameters of the proposed model could be realized with the current experiments. 
A relevant for the possible realisation of our proposal could be, for example, the experimental system reported in reference~~\cite{purdy2013observation}, with a silicon nitride membrane inside of a Fabry-Perot optical cavity.  The parameters achieved were $\omega_{m}/2\pi=1.55$ MHz, $\kappa/2\pi=0.89$ MHz, $\gamma/2\pi=0.47$ Hz, single-photon optomechanical coupling rate $g_{0}/2\pi=16$ Hz, and the photon number $N_{a}=3.6\times 10^{8}$, such that the effective coupling rate was estimated as $g/2\pi=0.3$ MHz. In a similar system reported in reference~\cite{peterson2016laser}, the parameters values achieved were $\omega_{m}/2\pi=1.48$ MHz, $\kappa/2\pi=2.6$ MHz, $\gamma/2\pi=0.18$ Hz, and $g/2\pi\le 30$ kHz. Furthermore, this system was prepared in a dilution refrigerator where $n_{m}\sim 10^{3}$ and could be cooled down to the ground state with $n_{0}=0.2$.  Besides, our proposal could be realized in microwave optomechanics, where LC resonators are used~\cite{fink2016quantum,barzanjeh2017mechanical}. In those experiments, parameters achieved were $\omega_m/2\pi\approx 5$ MHz, $\kappa/2\pi\approx 2$MHz, $\gamma/2\pi\approx 4$ Hz, $g_{0}/2\pi\approx 13\sim 34$ Hz and the photon number $N_{a}\approx 2350 \sim 4\times 10^{7}$ such that $g/2\pi=1.5 \sim 200$kHz. The LC circuit at milliKelvin temperatures in a dilution refrigerator can be cooled down to an average phonon occupancy of $n_{m}=0.32$. The linear coupling strength $J$  tuneable in a large range from $J\approx0.2$ MHz to $J\approx\omega_{m}\, (\omega_{m}/2\pi \approx 5\, {\rm MHz})$ have been reported~\cite{sankey2010strong,heinrich2011dynamics}. We point out that the computed theoretical results will be graphically presented for the feasible parameters chosen from the above mentioned experiments.

We now proceed to evaluate the populations of the modes and correlations between them.
We assume that $G>0$ that $G$ is the gain associated with the system. The input noise to the mechanical mode will be amplified during the evolution.
Therefore, we shall concentrate on short-time effects.
Using the standard cavity input-output relation $a^{\rm out}=a^{\rm in} + \sqrt{2\kappa}a$, we define annihilation operators of normalized temporal field modes~\cite{hofer2011quantum,he2014einstein}:
\begin{eqnarray}
A_{w}^{\rm in} &= &e^{-i\phi}\sqrt{\frac{2G}{1-e^{-2G\tau}}}\int_{0}^{\tau}dt\, a_{w}^{\rm in}(t)e^{-(G-i\delta)t}, \nonumber\\ 
A_{u}^{\rm in} &=& e^{i\phi}\sqrt{\frac{2G}{1-e^{-2G\tau}}}\int_{0}^{\tau}dt\, a_{u}^{\rm in}(t)e^{-(G-i\delta)t} ,\nonumber\\
A_{w}^{\rm out} &=& e^{i\phi}\sqrt{\frac{2G}{e^{2G\tau}-1}}\int_{0}^{\tau}dt\, a_{w}^{\rm out}(t)e^{(G+i\delta)t}, \nonumber\\
 A_{u}^{\rm out} &= &e^{-i\phi}\sqrt{\frac{2G}{e^{2G\tau}-1}}\int_{0}^{\tau}dt\, a_{u}^{\rm out}(t)e^{(G+i\delta)t} ,\nonumber\\
A^{\rm in}_m &=& c(0) ,\   A^{\rm out}_m = c(\tau) e^{-i\delta \tau} ,\label{II14}
\end{eqnarray}
where $\tau$ is the interaction time with the laser pulses.
It is easily verified that the operators obey the canonical commutation relations $[A_i,A_i^\dagger]=1$ ($i=u,w,m$). 

In what follows, we assume that the input fields to the field modes are in the ordinary vacuum state, the mechanical mode is initially in a thermal state with the mean number of phonons $n_{0}$. The mode is subjected to the damping $\gamma$ and to the Brownian noise $n_{m}$. Under these assumptions, we find the populations of the output modes to be
\begin{eqnarray}
\langle (A_{w}^{\rm out})^{\dag}A_{w}^{\rm out}\rangle &=& \frac{G_w}{G}\Upsilon(r) ,\quad  \langle (A_{u}^{\rm out})^{\dag}A_{u}^{\rm out}\rangle = \frac{G_u}{G}\Upsilon(r) ,\nonumber\\
\langle (A_{m}^{\rm out})^{\dag}A_{m}^{\rm out}\rangle &=& n_{0} + \Gamma(r) ,\label{II15}
\end{eqnarray}
and first-order correlation functions
\begin{eqnarray}
\langle (A_{w}^{\rm out})^{\dag}A_{u}^{\rm out}\rangle e^{i(\phi_{g_{w}}-\phi_{g_{u}})} = 
\langle (A_{u}^{\rm out})^{\dag}A_{w}^{\rm out}\rangle e^{-i(\phi_{g_{w}}-\phi_{g_{u}})}= \frac{\sqrt{G_wG_u}}{G}\Upsilon(r) ,\nonumber\\
\langle A_{m}^{\rm out}A_{w}^{\rm out} \rangle e^{-i\phi_{g_{w}}}= -i\sqrt{\frac{G_w}{G}} \Lambda(r) ,\ \langle A_{m}^{\rm out}A_{u}^{\rm out} \rangle e^{-i\phi_{g_{u}}} = -i\sqrt{\frac{G_u}{G}} \Lambda(r) ,\nonumber\\
\langle (A_{w}^{\rm out})^{\dag}(A_{u}^{\rm out})^{\dag}\rangle = \langle A_{w}^{\rm out}A_{u}^{\rm out}\rangle = \langle (A_{m}^{\rm out})^{\dag}A_{w}^{\rm out} \rangle = \langle (A_{m}^{\rm out})^{\dag}A_{u}^{\rm out} \rangle = 0 , \label{II16}
\end{eqnarray}
where  
\begin{eqnarray}
\Gamma(r) &=& \left(e^{2r}-1\right)\left[(n_0+1)+\frac{\gamma}{G}(n_m+1)\right] ,\nonumber\\
\Upsilon(r) &=& (n_{0}+1)\left(e^{2r}-1\right)+2\frac{\gamma}{G} (n_{m}+1)\frac{e^{2r} (\sinh2r - 2r)}{e^{2r}-1} ,\nonumber\\
\Lambda(r) &=& e^r\sqrt{e^{2r}-1}\left[n_0+1+\frac{\gamma}{G}(n_m+1)\left(1-\frac{2r}{e^{2r}-1}\right)\right] .\label{II17}
\end{eqnarray}
Here, $r=G\tau$ represents an effective squeezing parameter, $\phi_{g_{w}}$ and $\phi_{g_{u}}$ are phase angles associated with the coupling strengths of the superposition modes, $g_{w}=|g_{w}|\exp(i\phi_{g_{w}})$ and $ g_{u}=|g_{u}|\exp(i\phi_{g_{u}})$, respectively. 

Two remarks must be made about the solutions given by equations~(\ref{II15}) and (\ref{II16}). Firstly, there is no first-order coherence between the mechanical and superposition modes, but there is a nonzero anomalous coherence between these modes. Inversely, the first-order coherence is observed between the superposition modes but there is no anomalous coherence between these modes. This means that the first-order coherence between modes rules out the anomalous coherence between these modes and vice versa, the anomalous coherences between modes results in no first-order coherence. Secondly, the phases $\phi_{g_{w}}$ and $\phi_{g_{u}}$ do not affect measurable quantities such as populations of the modes, the first-order coherence between the modes or variances of the quadrature components of the modes. However, these measurable quantities can still be phase dependent since the absolute values $|g_{w}|^{2}$ and $|g_{u}|^{2}$ are not constant parameters. In the presence of the coupling $J$ these parameters turn out to be dependent on the phases of $g_{1}$ and $g_{2}$. It is easy to see. Writing $g_{1}=|g_{1}|\exp(i\phi_{g_1})$ and $g_{2}=|g_{2}|\exp(i\phi_{g_2})$, the phase-sensitive contribution to $|g_{w}|^{2}$ and $|g_{u}|^{2}$ can be identified, and the parameters $|g_{w}|^{2}$ and $|g_{u}|^{2}$ may be expressed as
\begin{eqnarray}
|g_{w}|^{2}\rightarrow |g_{w}(\psi)|^{2} &=& |g_{1}|^{2}\cos^{2}\theta +|g_{2}|^{2}\sin^{2}\theta + |g_{1}||g_{2}|\sin2\theta\cos2\psi ,\nonumber\\
|g_{u}|^{2}\rightarrow |g_{u}(\psi)|^{2} &=& |g_{1}|^{2}\sin^{2}\theta +|g_{2}|^{2}\cos^{2}\theta - |g_{1}||g_{2}|\sin2\theta\cos2\psi .\label{II18}
\end{eqnarray}
where $2\psi=\phi_{g_1}-\phi_{g_2}$ is the relative phase of the $g_{1}$ and $g_{2}$ coupling strengths. Clearly, the parameters $|g_{w}|$ and $|g_{u}|$ vary periodically with $2\psi$, but only if $\sin2\theta \neq 0$. For $J\neq 0$ we generally have $\sin2\theta\neq 0$, and then phase dependent effects are to be expected. On the other hand, for $J=0$, we have $\sin2\theta =0$ and then there is no phase dependence. 
Referring to equation~(\ref{II3}) the phases $\phi_{g_1}$ and $\phi_{g_2}$ can be controlled through the phases of the driving lasers and a constant phases difference $2\psi$ can be established. In an experimental situation, we would envisage a single laser providing the pump for both modes. 

We may summarize that the presence of the linear coupling $J$ between the field modes results in the phase dependence of the populations of the modes and coherences between them. In other words, the dependence of the solutions (\ref{II16}) on the relative phase $2\psi$ results from the presence of the three-mode loop in the coupling between the modes, as illustrated in figure~\ref{fig:scheme}(b).  This close loop coupling leads to interesting new effects, which will be discussed in details in the following section.

\section{Phase control of mutual coherence and populations of the optical fields}\label{sec.3}

We now turn to analyse the effect of the correlation between the superposition modes on the correlation and mutual coherence of the output field modes $A_1^{\rm out}$ and $A_2^{\rm out}$. 
To do this, we invert the transformations (\ref{II7}) and find
\begin{eqnarray}
a_{1} = a_{w}\cos\theta + a_{u}\sin\theta  ,\qquad  a_{2} = a_{w}\sin\theta - a_{u}\cos\theta .\label{III1}
\end{eqnarray}
Then, defining annihilation operators of the normalized temporal output modes 
\begin{eqnarray}
A_{j}^{\rm out} = \sqrt{\frac{2G}{e^{2G\tau}-1}}\int_{0}^{\tau}dt\, a_{j}^{\rm out}(t)e^{(G+i\delta)t} ,\quad j=1,2 .
\end{eqnarray}
we readily find that the relationship between $A_{1}^{\rm out}$, $A_{2}^{\rm out}$ and $A_{w}^{\rm out}$, $A_{u}^{\rm out}$ is 
\begin{eqnarray}
A_{1}^{\rm out} &=& A_{w}^{\rm out}e^{-i\phi}\cos\theta + A_{u}^{\rm out}e^{i\phi}\sin\theta ,\nonumber\\
A_{2}^{\rm out} &=& A_{w}^{\rm out}e^{-i\phi}\sin\theta - A_{u}^{\rm out}e^{i\phi}\cos\theta .\label{III3}
\end{eqnarray}

Then the populations of the modes and correlation functions are found to be
\begin{eqnarray}
	\langle (A_j^{\rm out})^{\dag}A_j^{\rm out}\rangle =  \frac{\kappa \Upsilon(r)}{G(\kappa^{2} +w^{2})}|{\cal A}_{j}(\psi)|^{2} , \quad  j=1,2, \nonumber\\
	\langle (A_1^{\rm out})^{\dag}A_2^{\rm out}\rangle e^{i(\phi_{{\cal A}_{1}}-\phi_{{\cal A}_{2}})} =\frac{\kappa \Upsilon(r)}{G(\kappa^{2} +w^{2})}|{\cal A}_{1}(\psi)||{\cal A}_{2}(\psi)| , \nonumber\\
	\langle A_m^{\rm out}A_j^{\rm out}\rangle e^{-i\phi_{{\cal A}_{j}}} = -i\sqrt{\frac{\kappa}{G(\kappa^2+w^2)}} \Lambda(r)\, |{\cal A}_{j}(\psi)| ,\nonumber\\
	\langle (A_{1}^{\rm out})^{\dag}(A_{2}^{\rm out})^{\dag}\rangle = \langle A_{1}^{\rm out}A_{2}^{\rm out}\rangle = 0,
\,        \langle (A_{m}^{\rm out})^{\dag}A_{j}^{\rm out} \rangle = 0 ,\label{III4} 
\end{eqnarray}
where  
\begin{eqnarray}
|{\cal A}_{j}(\psi)|^{2} &=& |g_{1}|^{2} +(-1)^j \left(|g_{1}|^{2} -|g_{2}|^{2}\right)\sin^{2}2\theta\sin^{2}\phi \nonumber\\
&-&(-1)^j |g_{1}||g_{2}|\left(\sin4\theta\sin^{2}\phi\cos2\psi - \sin2\theta\sin2\phi\sin2\psi\right) ,
\label{III5}
\end{eqnarray}
and we have extracted the phases $\phi_{{\cal A}_{j}}$ of the complex amplitude ${\cal A}_{j}(\psi) = |{\cal A}_{j}(\psi)|\exp(i\phi_{{\cal A}_{j}})$. The solutions (\ref{III4}) are of a simple form with the terms ${\cal A}_{1}(\psi)$ and ${\cal A}_{2}(\psi)$ representing the interference effects, while the terms $\Upsilon(r)$ and $\Lambda(r)$ give the time-dependent effect of the driving laser pulses. It is interesting to note that the interference terms factorize from the temporal terms that the interference effects are independent of the duration of the laser pulses. The solutions (\ref{III4}) have the same type correlation properties as in the case of the superposition modes, equations~(\ref{II15}) and (\ref{II16}). However, the dependence on the phases $\theta$ and $\psi$ is more complicated than that for the superposition modes and, in addition, it involves the phase $\phi$.

It should be noted here that for a given $r$ the total population of the modes is constant, independent of the phases $\theta$, $\phi$, and $\psi$,
\begin{eqnarray}
\langle (A_1^{\rm out})^{\dag}A_1^{\rm out}\rangle + \langle (A_2^{\rm out})^{\dag}A_2^{\rm out}\rangle &=& \langle (A_u^{\rm out})^{\dag}A_u^{\rm out}\rangle + \langle (A_w^{\rm out})^{\dag}A_w^{\rm out}\rangle \nonumber\\
&=& \frac{\kappa\left(|g_{1}|^{2}+|g_{2}|^{2}\right)}{G(\kappa^{2} +w^{2})}\Upsilon(r) .
\end{eqnarray}
In other words, the variation of the populations with the phase $\psi$ is due to the transfer of the population from one mode to the other not due to the generation or losses of photons.

\subsection{Mutual coherence and population transfer}

Let us discuss properties of the solutions (\ref{III4}) for some special cases. We are particularly interested in the properties of the mutual coherence between the output field modes.
Firstly, we note that for $J=0$ we have $\sin\theta=0, \cos\theta=1$, and then the coherence function $\langle (A_1^{\rm out})^{\dag}A_2^{\rm out}\rangle$ simplifies to
\begin{equation}
\langle (A_{1}^{{\rm out}})^\dag A_{2}^{\rm out}\rangle = \frac{\kappa \Upsilon(r)}{G(\kappa^{2} +\Delta^{2})}|g_{1}||g_{2}|e^{-i(\phi_{{\cal A}_{1}}-\phi_{{\cal A}_{2}})} .\label{III6}
\end{equation}
Clearly, in the absence of the coupling between the field modes $(J=0)$, the mutual coherence $|\langle (A_{1}^{{\rm out}})^\dag A_{2}^{\rm out}\rangle|$ is independent of the phases.

When the modes are degenerate $(\Delta=0)$, we have $\sin^{2}\theta =\cos^{2}\theta =1/2$ and then we obtain
\begin{eqnarray}
|\langle (A_{1}^{{\rm out}})^\dag A_{2}^{\rm out}\rangle| &=& \frac{\kappa \Upsilon(r)}{G(\kappa^{2} +J^{2})}\{|g_{1}|^{2}|g_{2}|^{2}\cos^{2}2\psi \nonumber\\
&+&[\frac{1}{2}(|g_{1}|^{2} -|g_{2}|^{2})\sin2\phi + |g_{1}||g_{2}|\cos2\phi\sin2\psi]^{2}\}^{\frac{1}{2}}  .\label{III7}
\end{eqnarray}

Further simplification to the case of the symmetric coupling $|g_{1}|=|g_{2}|=g$ gives
\begin{eqnarray}
|\langle (A_{1}^{{\rm out}})^\dag A_{2}^{\rm out}\rangle| &=& \frac{\kappa g^{2}\Upsilon(r)}{G(\kappa^{2} +J^{2})} \sqrt{1-\sin^{2}2\phi\sin^{2}2\psi} .\label{III8}
\end{eqnarray}
It follows that $|\langle (A_{1}^{{\rm out}})^\dag A_{2}^{\rm out}\rangle|$ vanishes only for $\sin^{2}2\phi=1$, i.e., for $\phi =\pi/4$ corresponding to $J=\kappa$. 
For $J\neq \kappa$, the mutual coherence never vanishes. 

It is interesting to noted that $\langle (A_{1}^{{\rm out}})^\dag A_{2}^{\rm out}\rangle$ can vanish even for an asymmetric coupling $|g_{1}|\neq |g_{2}|$. The expression~(\ref{III7}) is a sum of two positive numbers and therefore it would vanish when simultaneously both numbers are equal to zero. It is seen that both numbers are simultaneously equal to zero when $\cos2\psi= 0$, $\sin2\psi= \pm1$, and then $\tan2\phi = \mp 2|g_{1}||g_{2}|/(|g_{1}|^{2} -|g_{2}|^{2}) $. Thus, in the case of unequal coupling strengths the coherence function can vanish but only for a specific value of the ratio $J/\kappa$ that can satisfy this condition.

Apart from the mutual coherence, the populations of the superposition modes $A^{{\rm out}}_{w}$ and $A^{{\rm out}}_{u}$, as well as the modes $A^{{\rm out}}_{1}$ and $A^{{\rm out}}_{2}$ can depend on the phases. It is not difficult to see from equations~(\ref{II15}) and (\ref{II18}) that the populations of the superposition modes in the case $|g_{1}|\neq |g_{2}|$ never become zero as the phase $\psi$ is varied. For example, in the case $\psi=0$, the minimum of the population is simply proportional to $(|g_{1}|\sin\theta -|g_{2}|\cos\theta)^{2}/2$, and the maximum is proportional to $(|g_{1}|\cos\theta+|g_{2}|\sin\theta)^{2}/2$.
Only in the limit of $|g_{1}|=|g_{2}|=g$, the population may be completely transferred between the modes. In this case, the populations are given by
\begin{eqnarray}
\langle (A_{w}^{\rm out})^{\dag}A_{w}^{\rm out}\rangle &=& \frac{\kappa g^{2}\Upsilon(r) }{G(\kappa^{2} +w^{2})}\left(1+\sin2\theta\cos2\psi\right) ,\nonumber\\
\langle (A_{u}^{\rm out})^{\dag}A_{u}^{\rm out}\rangle &=& \frac{\kappa g^{2} \Upsilon(r) }{G(\kappa^{2} +w^{2})}\left(1-\sin2\theta\cos2\psi\right)  .\label{III10}
\end{eqnarray}
We see that in the absence of the linear coupling $(\sin2\theta =0)$ the modes are equally populated and the populations are independent of the phase $\psi$. The presence of the coupling $J$ clearly results in the populations becoming dependent on the phase $\psi$ and thus to allow the transfer of the population between the modes. However, the total transfer of the population takes place only when $\sin2\theta =1$, i.e., in the degenerate case of $\Delta =0$. Otherwise the populations can vary with the phase $\psi$ but cannot be totally transferred between the modes. For $\theta=\pi/4$ and $\psi = n\pi,\, (n=0,1,2,\ldots)$, we see from equation~(\ref{III10}) that the mode $u$ is unpopulated for all values of $r$, whereas the mode $w$ is maximally populated. On the other hand, for $\psi =(n+1/2)\pi$ this relationship is reversed and $u$ is the mode which is maximally populated. 

Consider now the populations of the output field modes, $A_{1}^{\rm out}$ and $A_{2}^{\rm out}$. From equations~(\ref{III4}) and (\ref{III5}) we see that apart from the dependence on the phase angles $\theta$ and $\psi$, the populations depend on the phase angle $\phi$. This may lead to some limits in the transfer of the populations not present for the superposition modes. It is easy to see from equation~(\ref{III5}) that for $J=0$ at which $\sin2\theta=0, \sin4\theta=0$, the populations are independent of the phases $\phi$ and $\psi$. On the other hand, for $\Delta=0$ and $J\neq 0$, we have $\sin2\theta=1, \sin4\theta=0$ and then 
\begin{eqnarray}
\langle (A_{j}^{\rm out})^{\dag}A_{j}^{\rm out}\rangle &=& \frac{\kappa \Upsilon(r)}{G(\kappa^{2} +J^{2})} \{ \frac{1}{2}(|g_{1}|^{2} +|g_{2}|^{2}) -(-1)^j\frac{1}{2}(|g_{1}|^{2} -|g_{2}|^{2})\cos2\phi
\nonumber\\
&&+(-1)^j |g_{1}||g_{2}|\sin2\phi\sin2\psi \}, \, (j=1,2)
\label{III11}
\end{eqnarray}
In this case the populations depend on the phases and are also strongly dependent on the relationship between the coupling constants $|g_{1}|$ and $|g_{2}|$. The involvement of the phase $\phi$ allows to achieve the complete transfer of the population between the states even if $|g_{1}|\neq |g_{2}|$. 

In particular, for the symmetric coupling $|g_{1}|=|g_{2}|=g$, we get
\begin{eqnarray}
\langle (A_{j}^{\rm out})^{\dag}A_{j}^{\rm out}\rangle &=& \frac{\kappa g^{2}\Upsilon(r)}{G(\kappa^{2} +J^{2})}\left[1+(-1)^j\sin2\phi\sin2\psi\right].
\label{III12}
\end{eqnarray}
Viewed as a function of $\psi$, the population can be transferred between the modes reversibly. However, there is a notable limit in the transfer of the populations provided by phase angle~$\phi$. Only for $\phi = \pi/4$ the population can be totally transferred from one mode to the other.

\subsection{Degree of coherence and visibility}

We have already seen that the coherence functions $\langle (A_{w}^{{\rm out}})^\dag A_{u}^{\rm out}\rangle$ and $\langle (A_{1}^{{\rm out}})^\dag A_{2}^{\rm out}\rangle$ vary with the phase angles $\theta$, $\phi$ and $\psi$, and may vanish for particular choices of the phases. However, if we calculate the the first-order coherence $\gamma_{ij}^{(1)}$ between the output modes, defined~as
\begin{equation}
\gamma_{jk}^{(1)} = \frac{|\langle (A_{j}^{{\rm out}})^\dag A_{k}^{\rm out}\rangle|}{\sqrt{\langle (A_{j}^{\rm out})^{\dag}A_{j}^{\rm out}\rangle\langle (A_{k}^{\rm out})^{\dag}A_{k}^{\rm out}\rangle}} ,\label{III13}
\end{equation}
we find that the first-order coherences $\gamma_{wu}^{(1)}$ and $\gamma_{12}^{(1)}$ are always unity, i.e., $\gamma_{wu}^{(1)}=1$ and $\gamma_{12}^{(1)}=1$.
This means that the superposition modes as well as the output cavity modes are mutually perfectly coherent irrespective of the values of the system's parameters. In other words, the phases of the modes are locked together. As we have already mentioned, the correlations between the modes and the populations are not constant, they vary with the parameters. This is illustrated in figure~\ref{fig:2} which shows the variation of the first-order correlation function $|\langle (A_{1}^{{\rm out}})^\dag A_{2}^{\rm out}\rangle|$ and the population of the field modes, $\langle (A_{1}^{\rm out})^{\dag}A_{1}^{\rm out}\rangle$ and $\langle (A_{2}^{\rm out})^{\dag}A_{2}^{\rm out}\rangle$, with the phase angle $\psi$ for equal $(|g_{1}| = |g_{2}|)$ and unequal $(|g_{2}| > |g_{1}|)$ couplings of the field  modes to the mechanical mode. It is clearly seen that the first-order coherence function and the populations vary periodically with the phase $\psi$. The coherence function is greatest half-way between the zeros of the populations and there are zeros in the coherence for phases at which the population is completely transferred to one of the modes. While figure~\ref{fig:2}(a) shows that in the case of $|g_{1}| = |g_{2}|$ the population is periodically transferred from one mode to the other, figure~\ref{fig:2}(b) shows that in the case of $|g_{1}|\neq |g_{2}|$ the complete transfer of the population occurs only to mode with larger coupling strength. 
\begin{figure}[t]
\center{
\includegraphics[width=0.46\textwidth]{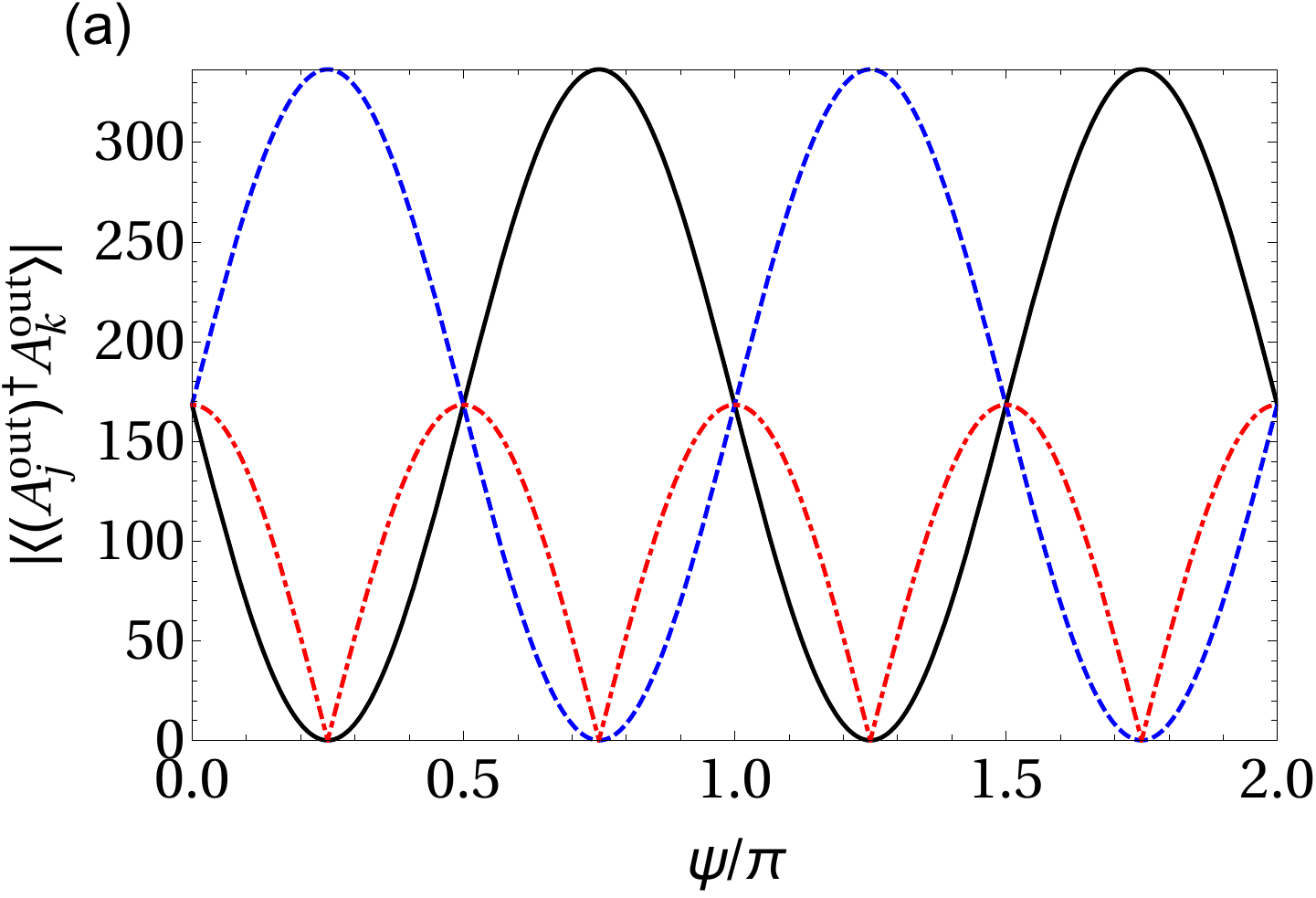}
\includegraphics[width=0.46\textwidth]{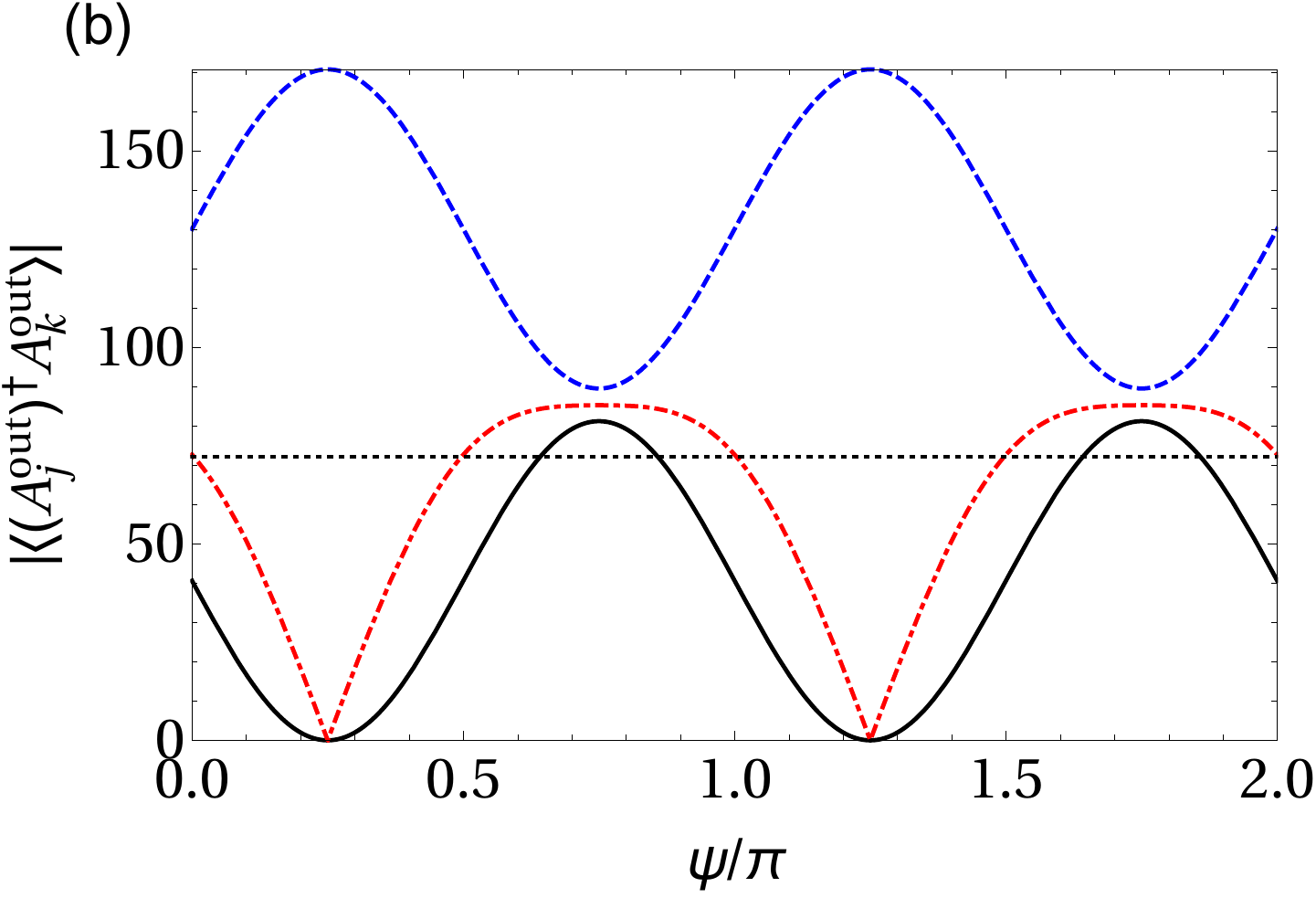}}
\caption{Variation of the first-order correlation function $|\langle (A_{1}^{{\rm out}})^\dag A_{2}^{\rm out}\rangle|$ (red dash-dotted line) and the population of the field modes $\langle (A_{1}^{\rm out})^{\dag}A_{1}^{\rm out}\rangle$ (black solid line) and $\langle (A_{2}^{\rm out})^{\dag}A_{2}^{\rm out}\rangle$ (blue dashed line) with the phase angle $\psi$ for $\Delta=0$, $\gamma =0.01g$, $n_{0}=n_{m}=200$, fixed $r=G\tau = 5$ and (a) equal coupling strengths $|g_1|=|g_2|=g$, $J=\kappa =10g$ where $g=0.1$ MHz, (b) unequal coupling strengths $|g_1|=g$, $|g_2|=2.5g$, $J=40g$, and $\kappa=100g$ where $g=0.01$ MHz. The horizontal black dotted line is the correlation function $|\langle (A_{1}^{{\rm out}})^\dag A_{2}^{\rm out}\rangle|$ for $J=0$.}
\label{fig:2}
\end{figure}

It is interesting that regardless of the distribution of the population between the modes the degree of coherence $\gamma_{jk}^{(1)}=1$. The constant phase relation between the modes is preserved even if one of the modes is not populated. This surprising behaviour is an example of induced coherence between the modes without induced emission, a phenomenon discussed and observed by several authors~\cite{zou1991induced,wang1991induced,horne1989two,wiseman2000induced,heuer2015induced,heuer2014phase,heuer2015complementarity,menzel2017entanglement}. In fact, the system considered here of two cavity modes coupled parametrically to the mechanical mode is analogous to the system of Wang, Zou and Mandel~\cite{zou1991induced,wang1991induced} composed of two coupled parametric downconverters in which the induced coherence without induced emission was demonstrated experimentally. In their experiment two downconverters were pumped by the same laser and arranged in cascade, that the idler field emitted from one of the downconverters was used as the input field of the other downconverter. Under this arrangement interference effects were observed between the signal fields of the two downconverters. It was further shown that in their system the degree of coherence $\gamma_{jk}^{(1)}=1$. In our system the two cavity modes are parametrically coupled to the mechanical mode and therefore can be treated as two downconverters. Since the modes are linearly coupled to each other through the coupling to the mechanical mode, they are locked in phase also and therefore exhibit perfect first-order coherence.

Although the modes $A_{1}^{{\rm out}}$ and $A_{2}^{\rm out}$ are mutually perfectly coherent, the visibility of the interference pattern can be zero that the modes can be completely distinguishable.   
The degree of visibility $\mathcal{V}$ of the interference pattern is related to distinguishability $\mathcal{D}$ through the expression $|\mathcal{V}|^{2} +|\mathcal{D}|^{2}\leq 1$ of complementarity. When $|\mathcal{V}| =1$ then $|\mathcal{D}|=0$ which means that the modes are indistinguishable. On the other hand, when $|\mathcal{V}|=0$ then $|\mathcal{D}|=1$ that the modes are perfectly distinguished.
The visibility $|\mathcal{V}|$ is given by
\begin{equation}
|\mathcal{V}| = \frac{2|\langle (A_1^{\rm out})^{\dagger} A_2^{\rm out}\rangle|} {\langle (A_1^{\rm out})^{ \dagger}A_1^{\rm out}\rangle + \langle (A_2^{\rm out})^{ \dagger} A_2^{\rm out}\rangle} 
= \frac{2|{\cal A}_{1}(\psi)||{\cal A}_{2}(\psi)|} {|{\cal A}_{1}(\psi)|^{2} +|{\cal A}_{2}(\psi)|^{2}} .\label{III15}
\end{equation}
Similarly, we may define a distinguishability
\begin{equation}
|\mathcal{D}| = \frac{|\langle (A_1^{\rm out})^{ \dagger}A_1^{\rm out}\rangle - \langle (A_2^{\rm out})^{ \dagger} A_2^{\rm out}\rangle|} {\langle (A_1^{\rm out})^{ \dagger}A_1^{\rm out}\rangle + \langle (A_2^{\rm out})^{ \dagger} A_2^{\rm out}\rangle} 
= \frac{|\left(|{\cal A}_{1}(\psi)|^{2} -|{\cal A}_{2}(\psi)|^{2}\right)|}{|{\cal A}_{1}(\psi)|^{2} +|{\cal A}_{2}(\psi)|^{2}} .\label{III16}
\end{equation}
Note that $|\mathcal{V}|^{2} +|\mathcal{D}|^{2}= 1$, which is a special case of the general expression $|\mathcal{V}|^{2} +|\mathcal{D}|^{2}\leq 1$.

If we take $\Delta =0$, we then find after substituting equation~(\ref{III5}) into equation~(\ref{III16}) that
\begin{equation}
|\mathcal{D}| = \frac{|\left(|g_{1}|^{2}-|g_{2}|^{2}\right)\cos2\phi -2|g_{1}||g_{2}|\sin2\phi\sin2\psi|}{|g_{1}|^{2} +|g_{2}|^{2}} .
\end{equation}
It is not difficult to check that the conditions for the modes to be perfectly indistinguishable $(\mathcal{D}=0)$ are different from the conditions for the modes to be perfectly distinguishable $(|\mathcal{D}|=1)$. For example, the modes are perfectly indistinguishable for the phase $\psi$ such that $\sin2\psi =\pm 1$ if $ \tan2\phi = \pm \left(|g_{1}|^{2}-|g_{2}|^{2}\right)/2|g_{1}||g_{2}| $. 
On the other hand, the modes can be perfectly distinguishable at the same phase $\psi$ only if $|g_{1}|=|g_{2}|$ and $\sin2\phi=1$, the latter happens when $J=\kappa$. Thus, for $J\neq \kappa$ the modes are always at least partly indistinguishable.

The phase dependent transfer of the population between the output cavity modes can be experimentally observed by measuring the coincidence rate $R_{12}$ between 
two detectors $D_{1}$ and $D_{2}$ adjusted to collect photons of the output modes $A_1^{\rm out}$ and $A_2^{\rm out}$, respectively.
The coincidence rate is proportional to the second order correlation function. In the case of the two output cavity fields, the rate is given by
\begin{equation}
R_{12} \sim \langle (A_1^{\rm out})^{\dagger} (A_2^{\rm out})^{\dagger}A_1^{\rm out}A_2^{\rm out}\rangle 
= 2\left[\frac{\kappa \Upsilon(r)}{G(\kappa^{2} +w^{2})}\right]^{2}|{\cal A}_{1}(\psi)|^{2}|{\cal A}_{2}(\psi)|^{2} .
\end{equation}
The choice of $\Delta =0$ and $|g_{1}|=|g_{2}|\equiv g$ leads to the result 
\begin{equation}
R_{12} = 2\left[\frac{\kappa g^{2}\Upsilon(r)}{G(\kappa^{2} +J^{2})}\right]^{2}\left(1-\sin^{2}2\phi\sin^{2}2\psi\right) .\label{III19}
\end{equation}
It is clear from equation~(\ref{III19}) that the coincidence rate exhibits a cosine modulation with the phase $\psi$. The relative depth of modulation is determined by $\sin^{2}2\phi$, which for $J=\kappa$ equals  unity. Under this condition, the coincidence rate vanishes when $2\psi = (n+1/2)\pi,\, (n=0,1,2,\ldots)$.
Comparing equation~(\ref{III19}) with equation~(\ref{III12}), we see that the coincidence rate vanishes for phases at which the population of either $A_1^{\rm out}$ or $A_{2}^{\rm out}$ is zero. In other words, $R_{12}=0$ signals the complete transfer of the population to one of the output cavity modes. 

In closing this section, we would like to point out that nonzero anomalous correlations are present between the mechanical and field modes, i.e., $\langle A_m^{\rm out}A_j^{\rm out}\rangle,\, (j=1,2,w,u)$ are all different from zero. These correlations are known to be responsible for entanglement between two modes. At the same time, all the first-order coherence functions $\langle (A_m^{\rm out})^{\dagger} A_j^{\rm out}\rangle$ are zero. It then follows that the mechanical and the field modes are mutually incoherent. Therefore, we may conclude that interference effects between two modes signal the complete separability of the modes and vice versa, entanglement signals the modes are mutually incoherent.  
In this sense, pairs of modes which are mutually coherent are clearly distinguishable from those which are entangled.

\section{Phase control of entanglement}\label{sec.4}

The presence of phase dependent populations of the modes and anomalous correlations between the mechanical and the field modes suggests that it could be possible to control and transfer of entanglement between these modes by manipulating the phase. To see the effects of the phase on entanglement between the modes, we examine the Duan-Simon inseparability criterion which involves linear combinations $X-X$, or $P-P$ or $X-P$ of the quadrature components of the modes.

The solutions given in equation~(\ref{II12}) show that the $X_{j}$ quadrature component is linearly related to the $P_{m}$ quadrature component, and vice versa, the $P_{j}$ quadrature component is linearly related to the $X_{m}$ quadrature component. Therefore, to quantify entanglement we will use the inseparability criterion for asymmetric $X - P$ combinations of the quadrature components of the output fields, which is determined by the separability parameter ($\hbar=1$)~\cite{duan2000inseparability,giovannetti2003characterizing,buono2012experimental,he2013einstein}
\begin{equation}
\Delta_{m,j} = \frac{\Delta^2(X^{\rm out}_{m}+h_{j} P^{\rm out}_{j})+\Delta^2(P^{\rm out}_{m}+h_{j} X^{\rm out}_{j})}{1+h_{j}^2} ,\quad j= 1,2,u,w ,\label{IV1}
\end{equation}
where $h_{j}$ is a weight factor which is chosen to minimize the variances. The optimal value of~$h_{j}$ which minimizes $\Delta_{i,j}$ is found using the variational method, 
\begin{equation}
h_{j} = \frac{\Delta^2P^{\rm out}_{j}-\Delta^2X^{\rm out}_{m}-\sqrt{(\Delta^2P^{\rm out}_{j}-\Delta^2X^{\rm out}_{m})^2 +4\langle X^{\rm out}_{m},P^{\rm out}_{j}\rangle^2}}{2\langle X^{\rm out}_{m},P^{\rm out}_{j}\rangle} .\label{IV2}
\end{equation}
Here the variances is defined as $\Delta^{2}O=\langle O^2 \rangle-\langle O \rangle^{2}$, and $\langle O_{1},O_{2} \rangle=(\langle O_{1}O_{2} \rangle+\langle O_{2}O_{1} \rangle)/2-\langle O_{1} \rangle\langle O_{2} \rangle$.
The modes $A^{\rm out}_{m}$ and $A^{\rm out}_{j}$ are said to be entangled if $\Delta_{m,j}<1$. The inequality $\Delta_{m,j}<1$ is necessary and sufficient condition to confirm entanglement between modes in Gaussian states under Gaussian measurements. The entanglement will be said to be maximal or perfect EPR state is created between the modes $A^{\rm out}_m$ and $A^{\rm out}_j$ if $\Delta_{m,j}$ can become zero. 

To determine the inseparability criterion we need the variances of the quadrature components of the modes and correlations between them. An obvious question is how to choose in our phase dependent system proper phase dependent quadratures which would correspond to the maximal reduction of fluctuations.  
The structure of the solutions given in equation~(\ref{II12}) suggests to define quadrature components of the optical fields that are in phase and out of phase with the coupling strengths of the modes to the mechanical mode. With this identification, we define the following quadrature components 
\begin{eqnarray}
X^{\rm out}_{m} = \frac{1}{\sqrt{2}}\left[A^{\rm out}_{m} +(A^{\rm out}_{m})^{\dag}\right] ,\quad P^{\rm out}_{m} = \frac{1}{\sqrt{2}i}\left[A^{\rm out}_{m} -(A^{\rm out}_{m})^{\dag}\right] ,\label{IV3}
\end{eqnarray}
for the output field of the mechanical mode, 
\begin{eqnarray}
X^{\rm out}_{j} &=& \frac{1}{\sqrt{2}}\!\left[A^{\rm out}_{j}e^{-i\phi_{g_{j}}} +(A^{\rm out}_{j})^{\dag}e^{i\phi_{g_{j}}}\right]  ,\nonumber\\
P^{\rm out}_{j} &=& \frac{1}{\sqrt{2}i}\!\left[A^{\rm out}_{j}e^{-i\phi_{g_{j}}} -(A^{\rm out}_{j})^{\dag}e^{i\phi_{g_{j}}}\right] , \label{IV4}
\end{eqnarray}
for the output field of the superposition modes $(j=u,w)$, and 
\begin{eqnarray}
X^{\rm out}_{j} &=& \frac{1}{\sqrt{2}}\!\left[A^{\rm out}_{j}e^{-i\phi_{{\cal A}_{j}}} +(A^{\rm out}_{j})^{\dag}e^{i\phi_{{\cal A}_{j}}}\!\right]  ,\nonumber\\
P^{\rm out}_{j} &=& \frac{1}{\sqrt{2}i}\!\left[A^{\rm out}_{j}e^{-i\phi_{{\cal A}_{j}}} -(A^{\rm out}_{j})^{\dag}e^{i\phi_{{\cal A}_{j}}}\!\right] ,\label{IV5}
\end{eqnarray}
for the output field of the cavity modes $(j=1,2)$.  
For the input field modes, we define the following quadrature components
\begin{eqnarray}
X^{\rm in}_{j} &=& \frac{1}{\sqrt{2}}\left[A^{\rm in}_{j} +(A^{\rm in}_{j})^{\dag}\right] ,\quad 
P^{\rm in}_{j} = \frac{1}{\sqrt{2}i}\left[A^{\rm in}_{j} -(A^{\rm in}_{j})^{\dag}\right] ,\ j=1,2,u,w, \nonumber\\
X^{\rm in}_{m} &=& \frac{1}{\sqrt{2}}\left[A^{\rm in}_{m} +(A^{\rm in}_{m})^{\dag}\right] ,\quad P^{\rm in}_{m} = \frac{1}{\sqrt{2}i}\left[A^{\rm in}_{m} -(A^{\rm in}_{m})^{\dag}\right] .\label{IV6}
\end{eqnarray}

For the field modes initially in the ordinary vacuum state the variances of the quadrature components of the input fields are $\Delta^2X^{\rm in}_{1}=\Delta^2P^{\rm in}_{1}=\Delta^2X^{\rm in}_{2}=\Delta^2P^{\rm in}_{2}=1/2$. For the mechanical mode initially in a thermal state, $\Delta^2X^{\rm in}_{m}=\Delta^2P^{\rm in}_{m} = n_{0}+1/2$. To determine variances of the output field quadratures and correlations between them, which are needed in equation~(\ref{IV1}), we make use of equations~(\ref{II16}) and (\ref{III4}), and find 
\begin{eqnarray}
\Delta^{2}X^{\rm out}_{m} = \Delta^{2}P^{\rm out}_{m} = n_{0} +\frac{1}{2} +\Gamma(r) ,\nonumber\\
\Delta^{2}X_{j}^{\rm out} = \Delta^{2}P_{j}^{\rm out} = \frac{1}{2} +\Upsilon(r)|{\cal U}_{j}(\psi)|^{2} ,\nonumber\\
\langle X_{m}^{\rm out},P_{j}^{\rm out}\rangle = \langle P_{m}^{\rm out},X_{j}^{\rm out}\rangle = -\Lambda(r)|{\cal U}_{j}(\psi)| ,\label{IV7}
\end{eqnarray}
where 
\begin{eqnarray}
|{\cal U}_{j}(\psi)| = \left\{
\begin{array}{c}
\sqrt{\frac{\kappa}{G(\kappa^2+w^2)}}|g_{j}(\psi)| ,\quad {\rm for}\quad   j=u,w,  \\
\sqrt{\frac{\kappa}{G(\kappa^2+w^2)}}|{\cal A}_{j}(\psi)| ,\quad {\rm for}\quad  j=1,2 .  
\end{array}
\right .\label{IV8a}
\end{eqnarray}
Note that the variances of the quadratures of the output fields are all greater than $1/2$. This means that the fluctuations in the output modes are not squeezed and therefore correlations between the modes are necessary to produce entanglement. 

To illustrate the analytic structure of the parameter $\Delta_{m,j}$, we consider two simpler but standard cases where one case is the minimized $\Delta_{m,j}$ by the optimal gain factor with the limit $\gamma/G \ll 1$~\cite{purdy2013observation} and the other case is $\Delta_{m,j}|_{h_j=1}$ for a fixed gain factor $h_j=1$. 

After making use of equation~(\ref{IV7}) in equation~(\ref{IV1}), we find that in the limit of $\gamma/G\ll 1$, the parameter $\Delta_{m,j}$ is given by
\begin{equation}
\Delta_{m,j} = \frac{h_{j}^{2}-1}{h_{j}^{2}+1} +2\frac{(n_{0}+1)e^{2r}}{h_{j}^{2}+1}\!\left\{1-h_{j}\sqrt{1-e^{-2r}}|{\cal U}_{j}(\psi)|\right\}^{2} .\label{IV8}
\end{equation}
It is clear from the form of the two terms on the right-hand side of equation~(\ref{IV8}) that the separability parameter $\Delta_{m,j}$ attains a minimum value, corresponding to maximum entanglement, when the term inside the curl brackets is equal to zero. It happens when $e^{-2r}\approx 0$ and $h_{j}|{\cal U}_{j}(\psi)|=1$, in which case $\Delta_{m,j} = (h_{j}^{2}-1)/(h_{j}^{2}+1)$. It follows that optimal  entanglement $(\Delta_{m,j}=0)$ could be observable in principle for the symmetric $(h_{j}=1)$ combination of the $X_{m}$ and $P_{j}$ quadrature components and sufficiently long laser pulses.

However, the question whether $\Delta_{m,j}<1$ and whether it is possible to achieve $\Delta_{m,j}=0$ depends strongly on the phase dependent factors $|{\cal U}_{j}(\psi)|$ and the optimal gain factor $h_j$. The variation of the parameters $\Delta_{m,j}$ with the relative phase $\psi$ for large $r$ is plotted in figure~\ref{fig:3}. The horizontal dashed-dotted line represents the degree of entanglement in the absence of the coupling $J$. In this case both field modes are equally entangled with the mechanical mode and the entanglement is independent of $\psi$. In fact, $\Delta_{m,j}$ can never be reduced to zero even for large $\tau$ due to the presence of initial thermal noise in the mechanical mode. In the presence of the coupling $J$, the separability parameters oscillate periodically with the phase. The amplitude of the oscillations is equal to one. The most interesting is that optimal entanglement $(\Delta_{m,j}=0)$, resulting in perfect EPR state, is achieved at $\sin2\psi =\pm 1$ for $j=1,2$ and $\cos2\psi = \pm 1$ for $j=w,u$. For example, the parameter $\Delta_{m,1}=0$ happens periodically for the phase satisfying $\psi = (n+3/4)\pi, \, (n=0,1,2,\ldots)$ where the optimal gain $h_{1}=1$ and the phase dependent factor $|{\cal U}_{1}(\psi)|=1$. Meanwhile, $\Delta_{m,2}=1$ at these values of the phase since there is no population in the mode $A^{\rm out}_{2}$. On the other hand, $\Delta_{m,w}=0$ happens for the phase $\psi$ satisfying $\psi = n\pi,\, (n=0,1,2,\ldots)$ where the optimal gain $h_{w}=1$ and $|{\cal U}_{w}(\psi)|=1$.  Meanwhile, $\Delta_{m,u}=1$ at these phases since there is no population in the mode $A^{\rm out}_{u}$. Note that the case $\Delta_{m,j}=0$ occurs for phases $\psi$ at which either $A^{\rm out}_{1}$ and $A^{\rm out}_{2}$ or $A^{\rm out}_{w}$ and $A^{\rm out}_{u}$ are distinguishable.

In addition, the system behaves deterministically that the entanglement can be periodically transferred between the modes by varying the relative phase $\psi$. This is of course a reflection of the fact that the field modes are perfectly coherent and the coherence is preserved independent of the redistribution of the population between the modes. Note that the location of the zeros of $\Delta_{m,j}$ corresponds to the zeros of the population of the other field mode (see figure~\ref{fig:2}). Thus, a constant phase relation between the modes allows for coherent transfer of the population and correlations between the modes.

\begin{figure}[h]
\centering
\includegraphics[width=0.46\textwidth]{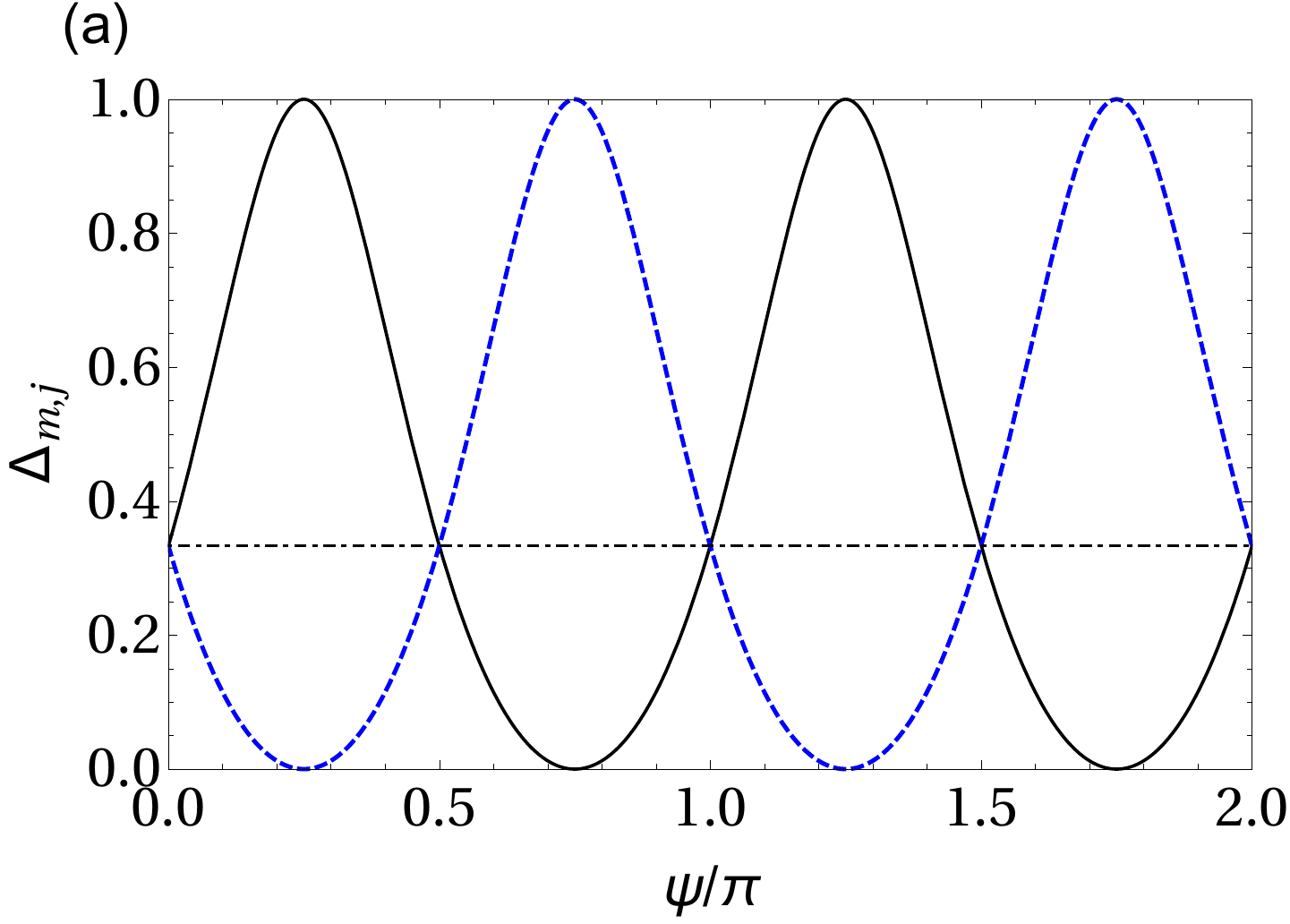}
\includegraphics[width=0.46\textwidth]{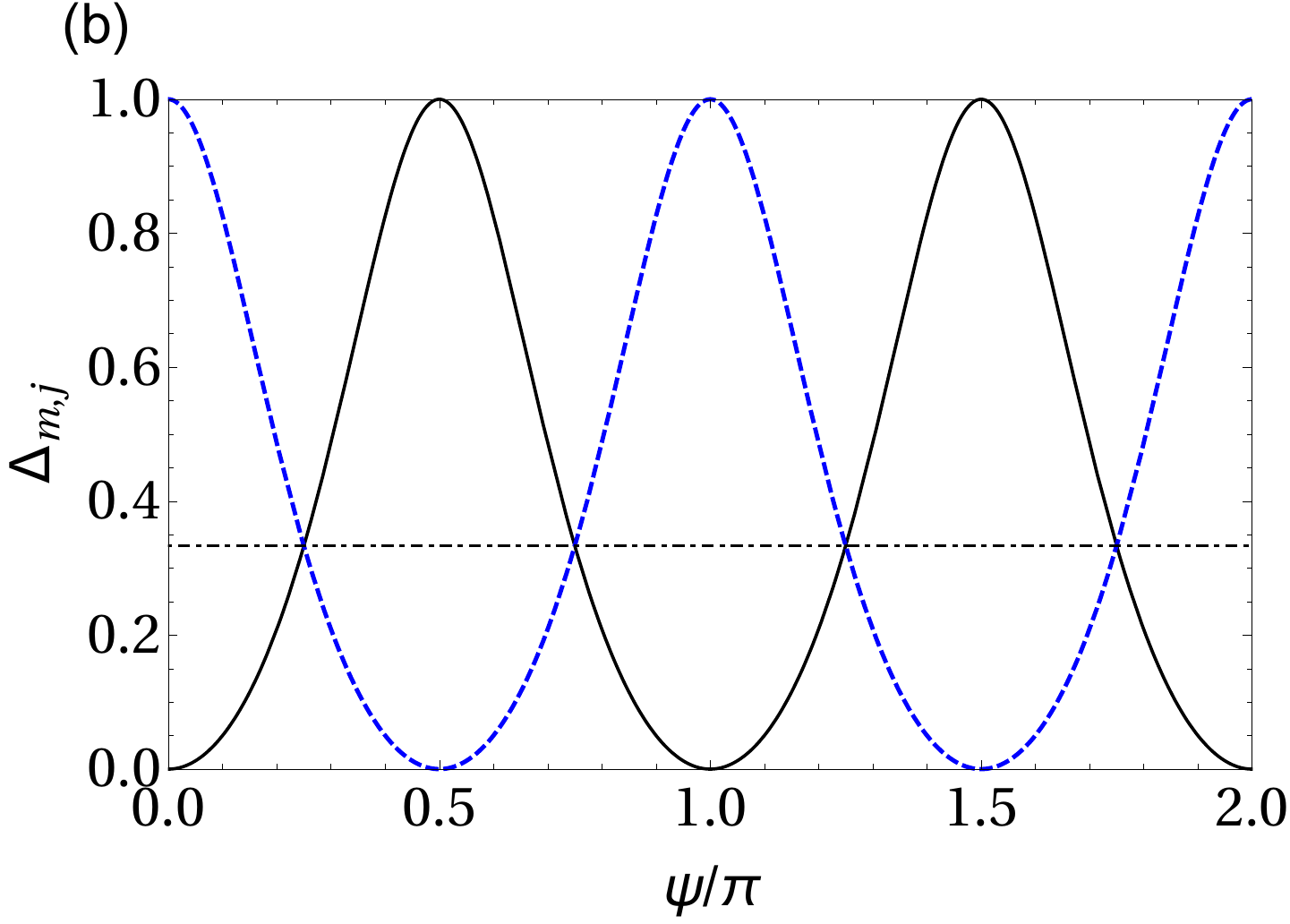}
\caption{The separability parameters $\Delta_{m,j}$ given in equation~(\ref{IV8}) plotted as a function of the phase $\psi$ for the optimal superposition of the quadrature components, $|g_1|=|g_2|\equiv g$, $J=\kappa=10g$, $\Delta=0$, $\gamma=0$, $n_{0}=n_{m}=200$, and a fixed $r=G\tau =5$. Here, $g=0.1$ MHz. In frame (a) the solid black line is for $\Delta_{m,1}$, dashed blue line is for $\Delta_{m,2}$.  In frame (b) the solid black line is for $\Delta_{m,w}$, dashed blue line is for $\Delta_{m,u}$. The horizontal dashed-dotted line in both figures represent the  results for~$J=0$.}
\label{fig:3} 
\end{figure}
\begin{figure}[h]
\centering
\includegraphics[width=0.46\textwidth]{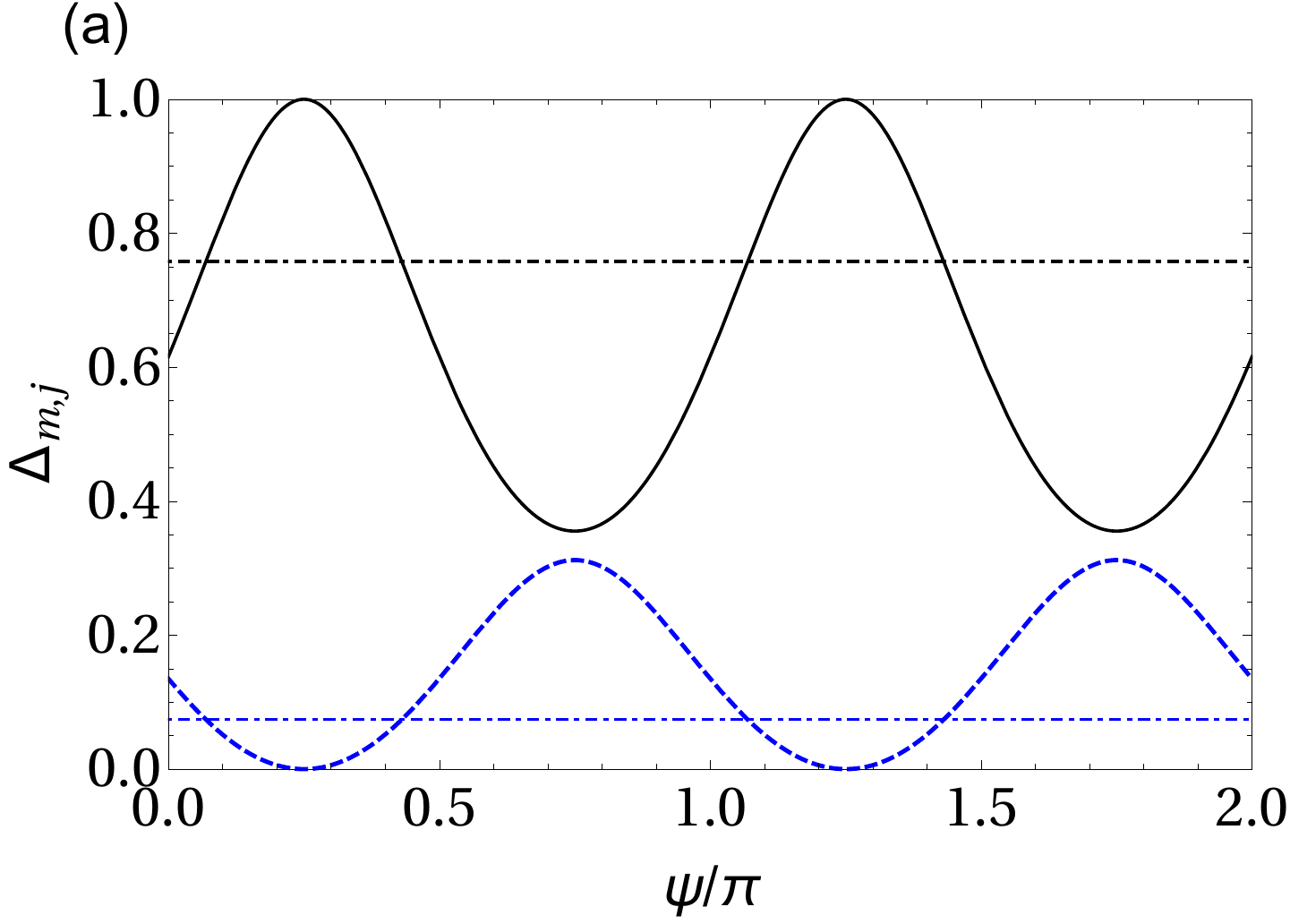}
\includegraphics[width=0.46\textwidth]{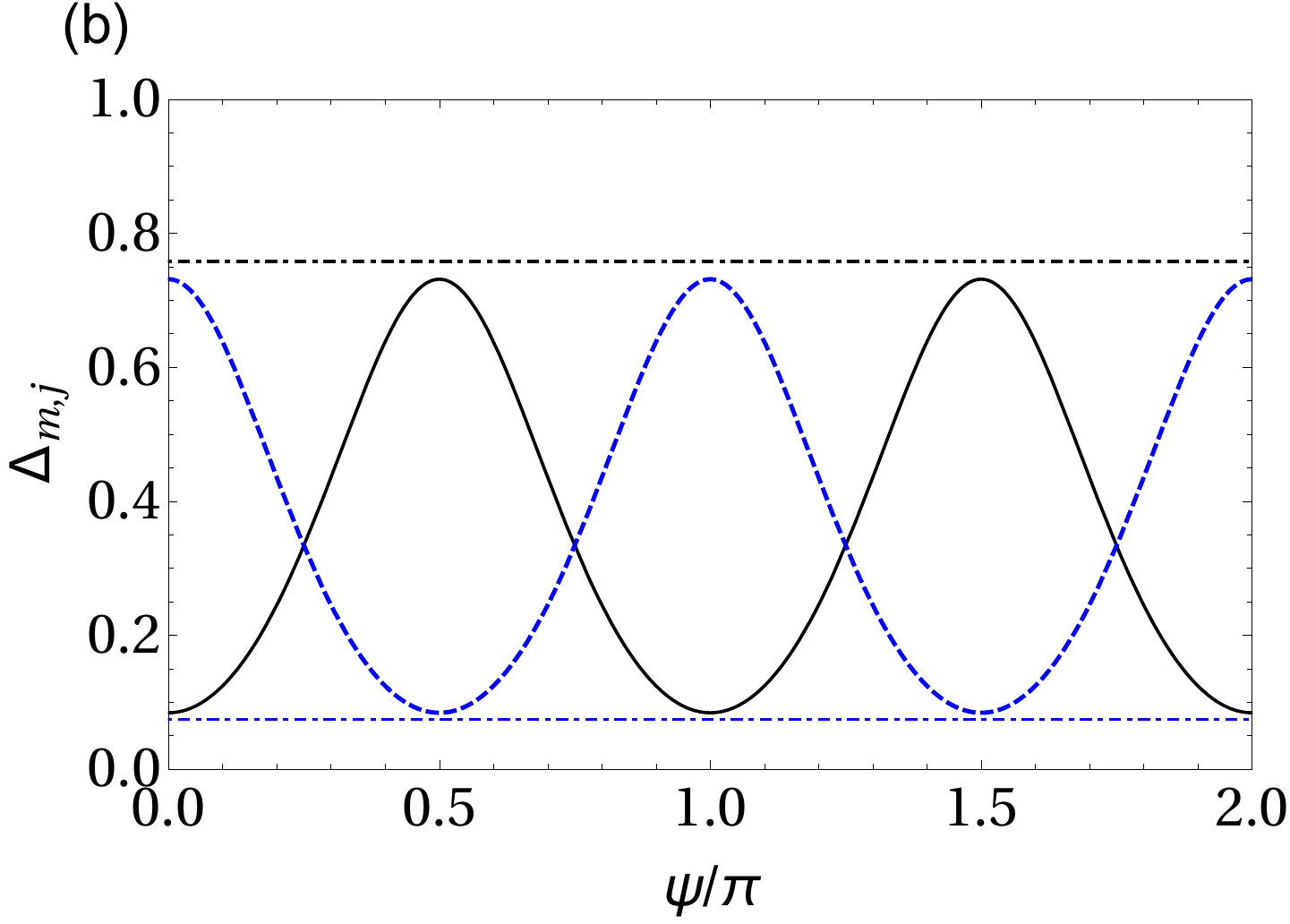}
\caption{The separability parameters $\Delta_{m,j}$ given in equation~(\ref{IV8}) plotted as a function of the phase $\psi$ for $|g_1|\equiv g$, $|g_{2}|=2.5g$, $\kappa=10g$, $J=4g$, $\Delta=0$, $\gamma=0$, $n_{0}=n_{m}=200$, and a fixed $r=G\tau=6.25$. Here $g=0.1$ MHz. In frame (a), the solid black line is for $\Delta_{m,1}$, dashed blue line is for $\Delta_{m,2}$, and the horizontal dashed-dotted lines represent these parameters for $J=0$. In frame (b) the solid black line is for $\Delta_{m,w}$, dashed blue line is for $\Delta_{m,u}$. The horizontal dashed-dotted lines represent the corresponding parameters for~$J=0$.}
\label{fig:4} 
\end{figure}

Figure~\ref{fig:4} shows the corresponding situation for unequal coupling strengths, $|g_{1}|\neq |g_{2}|$ with the degenerate modes $\Delta=0$ and $\gamma=0$. We see that the amplitudes of the periodic maxima and minima of $\Delta_{m,j}$ occur at the same phases as in figure~\ref{fig:3}, but are noticeable reduced in magnitude. Figure \ref{fig:4}a shows that the amplitude of $\Delta_{m,2}$ can be reduced to zero at $\psi = (n+1/4)\pi, \, (n=0,1,2,\ldots)$ where $h_2=1, \,|{\cal U}_{2}(\psi)|=1$, while $\Delta_{m,1}$ cannot due to the required condition cannot be satisfied. 
This is easy to understand if one refers to the fact that in the case $|g_{2}|>|g_{1}|$ only the population of the mode $A^{\rm out}_{1}$ can be completely transferred to the mode  $A^{\rm out}_{2}$. When the population is completely transferred $\Delta_{m,1}=1$ and simultaneously $\Delta_{m,2}=0$. Figure~\ref{fig:4}b shows that the behavior of the parameters $\Delta_{m,w}$ and $\Delta_{m,u}$ is much different than $\Delta_{m,1}$ and $\Delta_{m,2}$. The amplitudes of their oscillation are equal but not one anymore, and become very small for greatly unequal coupling strengths. Thus we conclude that the result of unequal coupling strengths is marked, always reducing the amplitudes of the variation of the separability parameters, and optimal entanglement occurs only between the mechanical mode and one of the cavity modes whose the coupling strength is larger than the other mode.

Now, let us illustrate the analytic expression of the second case where $h_{j}=1$, $\Delta=0\, (\theta=\pi/4)$, and $|g_{1}|=|g_{2}|$. By substituting for $|{\cal U}_{j}(\psi)|$ from equation~(\ref{III5}) we obtain an explicit solution for the phase dependence of the separability parameters $ (j=1,2)$
\begin{equation}
\Delta_{m,j}|_{h_j=1} = (n_{0}+1)e^{2r}\!\left\{1-\sqrt{\frac{1}{2}\left(1-e^{-2r}\right)\left[1+(-1)^{j}\sin2\phi\sin2\psi\right]}\right\}^{2}.\label{IV9}
\end{equation}
Similarly, by substituting for $|{\cal U}_{j}(\psi)|$ from equation~(\ref{II18}) we obtain an explicit solution for the phase dependence of the separability parameters  $ (j=w,u)$,
\begin{equation}
\Delta_{m,j}|_{h_j=1} = (n_{0}+1)e^{2r}\!\left\{1-\sqrt{\frac{1}{2}\left(1-e^{-2r}\right)\left(1\pm\cos2\psi\right)}\right\}^{2},\label{IV10}
\end{equation}
in which the upper sign ``$+$''  at $\cos2\psi$ is for $j=w$ and the lower sign ``$-$'' is for $j=u$. 

It is seen from equations~(\ref{IV9}) and (\ref{IV10}) that $\Delta_{m,j}|_{h_j=1}$ varies periodically with the phase $2\psi$. It is not difficult to verified that the periodicity is the same as the periodicity of the oscillation of the populations of the mode $j$. For a sufficiently large squeezing parameter $r$, such that $e^{-2r}\approx 0$, choosing $\sin2\phi=1$, one finds from equation~(\ref{IV9}) that the choice of the relative phase $\sin2\psi=\pm 1$ leads to a reduction of $\Delta_{m,1}|_{h_1=1}$ and $\Delta_{m,2}|_{h_2=1}$ to zero. Similarly, for a particular choice of $\psi$ at which $\cos2\psi=\pm 1$, the term inside the curl brackets of equation (\ref{IV10}) is equal to zero. Thus, $\Delta_{m,u}|_{h_u=1}$ and $\Delta_{m,w}|_{h_w=1}$ can be reduced to zero. It follows that perfect entanglement is achievable. Hence, with the presence of the linear coupling $J$, optimal entanglement can be created between the mechanical and one of the field modes. Note that the optimal entanglement in this case is same with the first case given in equation~(\ref{IV8}), where the optimal gain factor $h_j=1$ at same specific choices of phase for mode $j$. Moreover, we have also investigated the effects of phase fluctuations on the optimal entanglement by involving the phase noises of the two driving lasers $\varphi_{1}(t)$ and $\varphi_{2}(t)$ in the Hamiltonian, the details can be found in~\ref{sec.A1}. There we find that the optimal entanglement is reduced for a finite correlation time of the phase noise process. However, we can still achieve entanglement when the phase noise is in a reasonable size.

We have so far ignored the effect of the damping $\gamma$ of the mechanical mode as this has been secondary to our goal of demonstrating how entanglement varies with the phase. 
However, to see how the damping $\gamma$, if not ignored could affect the entanglement between the modes, we plot in figure~\ref{fig:5} the separability parameter $\Delta_{m,j}$ as a function of $r$ for several different values of $\gamma/G$. It is seen that as soon as $\gamma/G\neq 0$, there is no optimal entanglement, but the modes can be significantly entangled over entire range of $r$ when $\gamma(n+1)<G$. Even for $\gamma(n+1)>G$ the mode still can be entangled but the entanglement is restricted to small $r$. It is seen that for large $\gamma$, the  parameter $\Delta_{m,j}$ passes through a minimum, which occurs at small values of~$r$, and then tends to a nonzero value $\gamma(n+1)/G$ as $r$ increases. Notice that the value at which~$\Delta_{m,j}$ saturates is smaller than unity when $\gamma(n+1)<G$.

\begin{figure}[h]
\centering
\includegraphics[width=0.5\textwidth]{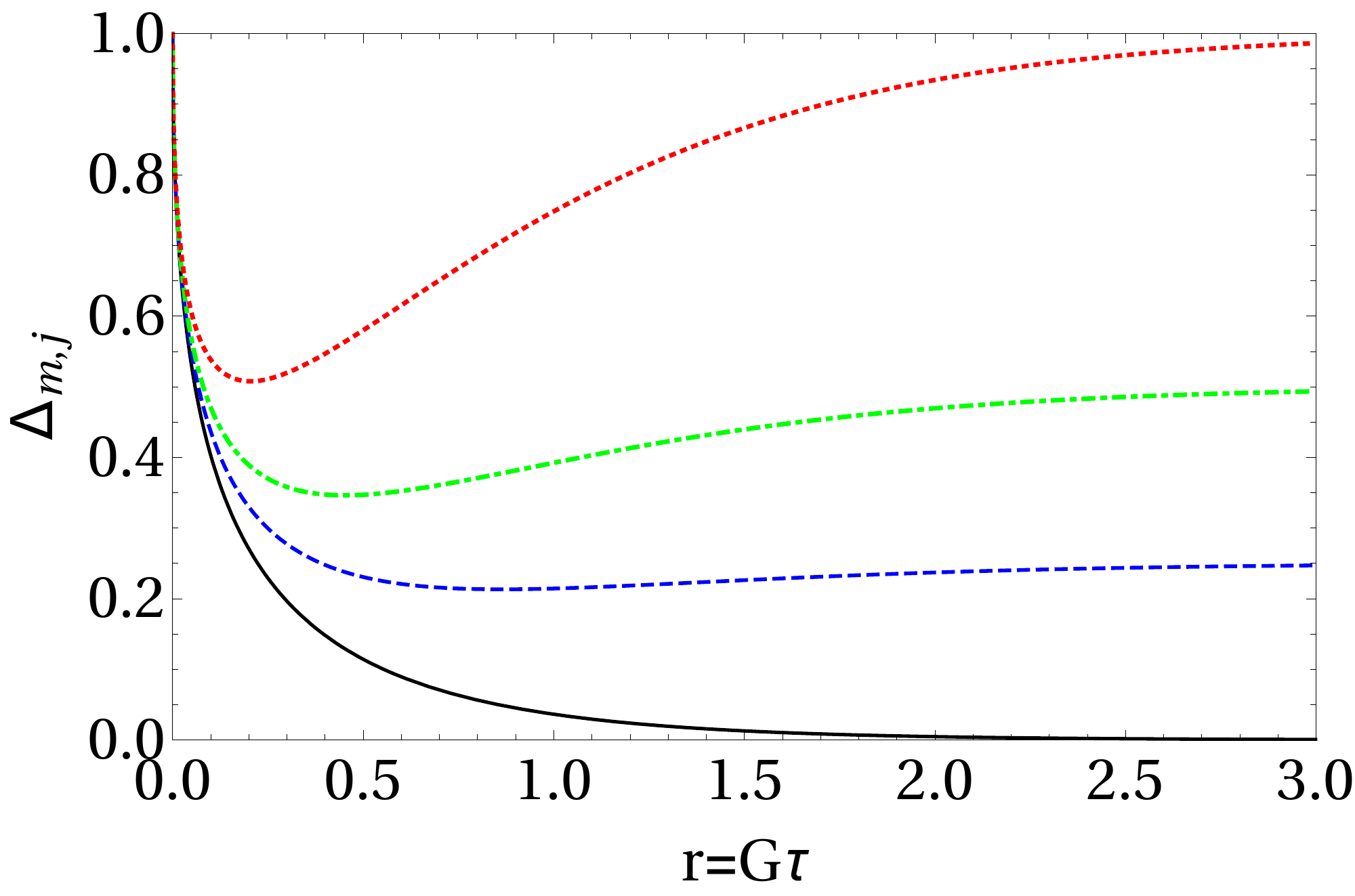} 
\caption{Separability parameter $\Delta_{m,j}$ plotted as a function of $r=G\tau$ for $\Delta=0$, $n\equiv n_{0}=n_{m}=0$, $|g_{1}|=|g_{2}|=0.1$ MHz, the phase $\psi$ such that $|{\cal U}_{j}(\psi)|=1$, and various damping rates $\gamma$: $\gamma =0$ (black solid line), $\gamma =0.25G$ (blue dashed line), $\gamma =0.5G$ (green dashed-dotted line), and $\gamma = G$ (red dotted line).}
\label{fig:5} 
\end{figure}

The behavior of $\Delta_{m,j}$ can be understood as follows. For $\Delta=0$, $|g_{1}|=|g_{2}|$, and the particular values of $\psi$ at which $|{\cal U}_{j}(\psi)|=1$, the separability parameter $\Delta_{m,j}$ can be expresses as the sum of two terms ($n\equiv n_0=n_m$)
\begin{eqnarray}
\Delta_{m,j} &=& (n+1)\left\{e^{2r}\left(1+\frac{\gamma}{G}\right)\left(1-\sqrt{1-e^{-2r}}\right)^{2}\right. \nonumber\\
&&\left. +\frac{\gamma}{G}\left[1-\frac{4r\left(1-\sqrt{1-e^{-2r}}\,\right)}{1-e^{-2r}}\right]\right\} .\label{IV11}
\end{eqnarray}
Notice that the first term on the right-hand side (RHS) of equation~(\ref{IV11}) exhibits the same variation with $r$ as in the absence of the damping. This term tends to zero as~$r$ increases. Therefore, the second term on RHS represents the major contribution of the damping. This term tends to reduce the entanglement but only for large $r$. It is easily found that in the case of small $r$, the second term can be approximated~by $-\left(1-2\sqrt{2r}\right)\gamma/G$,
from which we see that there is a threshold value of $r=1/8$ below which this term is negative. This has the effect that below threshold $(r<1/8)$ the parameter $\Delta_{m,j}$ is not much affected by the damping $\gamma$, the negative term compensates a part of the positive contribution of $\gamma$ to the amplitude of the first term. 
Above threshold $r>1/8$, the second term is positive resulting in an increased value of $\Delta_{m,j}$. Nevertheless, even in this case $\Delta_{m,j}$ can still be smaller than unity. It is easy to see, for large values of $r$, such that $e^{-2r}\approx 0$, the first term on RHS of equation~(\ref{IV11}) vanishes, but the second does not and takes a simple asymptotic value
$\Delta_{m,j} =\left(n+1\right) \gamma/G$. 
Thus, in the presence of the damping $\gamma$, the separability parameters saturate at a nonzero value, which is smaller than unity whenever $\gamma(n+1)<G$. Hence we may conclude  that~$\Delta_{m,j}$ is not affected much by the damping $\gamma$ when $r<1/8$. The damping $\gamma$ tends to affect $\Delta_{m,j}$ at large values of $r$ and causes the separability parameter to saturate at the value $\gamma(n+1)/G$.

\section{Phase control of quantum steering}\label{sec.5}

In Section~\ref{sec.4} we discussed the control and transfer of entanglement between the mechanical and field modes by manipulating the relative phase of their coupling strengths. 
In this section, we discuss how such phase sensitive coupling may be used to control quantum steering of the modes.  

Quantum steering takes place between entangled modes and provides the information as to how a given mode ``steers'' the other mode to be entangled~\cite{wiseman2007steering,jones2007entanglement}. The quantum steering of mode $i$ by mode $j$ can be identified by $E_{i|j}=\Delta_{inf,j}X_{i}\Delta_{inf,j}P_{i}<1/2$ ($\hbar=1$),  where the inferred quadrature variances are defined as $\Delta^2_{inf,j}X_i=\Delta^2(X_i+u_jO_j)$ and $\Delta^2_{inf,j}P_i=\Delta^2(P_i+u'_j O^{\prime}_j)$, $O_j$, $O^{\prime}_j$ are arbitrary observables of the system $j$, and $u_j$, $u'_j$ are gain factors. The quadrature components and the gain factor are selected such that they minimize the inference (conditional standard deviation) product.  In our case, $\Delta^{2}X_{j}=\Delta^{2}P_{j}$ as shown in equation~(\ref{IV7}), and we choose $O_{j}=P_{j}$ and $O^{\prime}_{j}=X_{j}$, the optimal gain factor is obtained as $u_{j}=u'_{j}=-\langle X_{i},P_{j} \rangle/\Delta^{2}P_{j}$ via variation method $\partial E_{i|j} / \partial u_j =0$.
As a result, we have
\begin{equation}
E_{i|j} =  \Delta^{2}X_{i} -\frac{\langle X_{i},P_{j}\rangle^{2}}{\Delta^{2}P_{j}}.\label{V1}
\end{equation}

We see from equation~(\ref{V1}) that a nonzero correlation $\langle X_{i},P_{j}\rangle$ is required to achieve the steering condition. In Section~\ref{sec.4}, we found that the nonzero correlations exist only between the mechanical mode and the output cavity modes as well as between the optomechanical mode and the output superposition modes.
Therefore, we will look at the steering of the mechanical mode by either one of the output cavity modes or one of the output superposition modes. Substituting the variances and correlations given in equation~(\ref{IV7}) into equation~(\ref{V1}), we can calculate the steering parameters analytically. In the limit of $\gamma \ll G$, we get a simple analytical expression of the steering parameter 
\begin{eqnarray}
E_{m|j} = \frac{1}{2}+\frac{(n_{0}+1)(e^{2r}-1)\left(1-2|{\cal U}_{j}(\psi)|^{2}\right)+n_{0}}{1+2(n_{0}+1)(e^{2r}-1)|{\cal U}_{j}(\psi)|^{2}} ,\label{V2}
\end{eqnarray}
where $|{\cal U}_{j}(\psi)|^{2}$ is given in equation~(\ref{IV8a}). We see that the mechanical mode can be steered in a similar manner by the two field modes $(j=1,2)$ and their superposition modes $(j=u,w)$. The phase dependent function $|{\cal U}_{j}(\psi)|^{2}$ plays the major role in the steering properties of the modes. If the relative phase $\psi$ is such that $|{\cal U}_{j}(\psi)|^{2}=1$, then the steering parameter $E_{m|j}$ falls below its vacuum level $1/2$ by a minimum value which dependent directly on $n_{0}$ and the degree of squeezing~$r$. If $n_{0}=0$ and if $|{\cal U}_{j}(\psi)|^{2}=1$, then quantum steering $(E_{m|j}<1/2)$ is seen to occur over the entire range of $r$. When $n_{0}\neq 0$, quantum steering is possible for $r>r_{th}$, where $r_{th} =\ln[(2n_{0}+1)/(n_{0}+1)]/2$. The minimum value of $E_{m|j}$ is reached periodically when $|{\cal U}_{j}(\psi)|^{2}=1$ and $r\gg 1$, in which case $E_{m|j}=0$ and the steering is perfect. Thus, we may speak about perfect EPR state between the mechanical mode $m$ and mode $j$. 

\begin{figure}[h]
\centering
\includegraphics[width=0.46\textwidth]{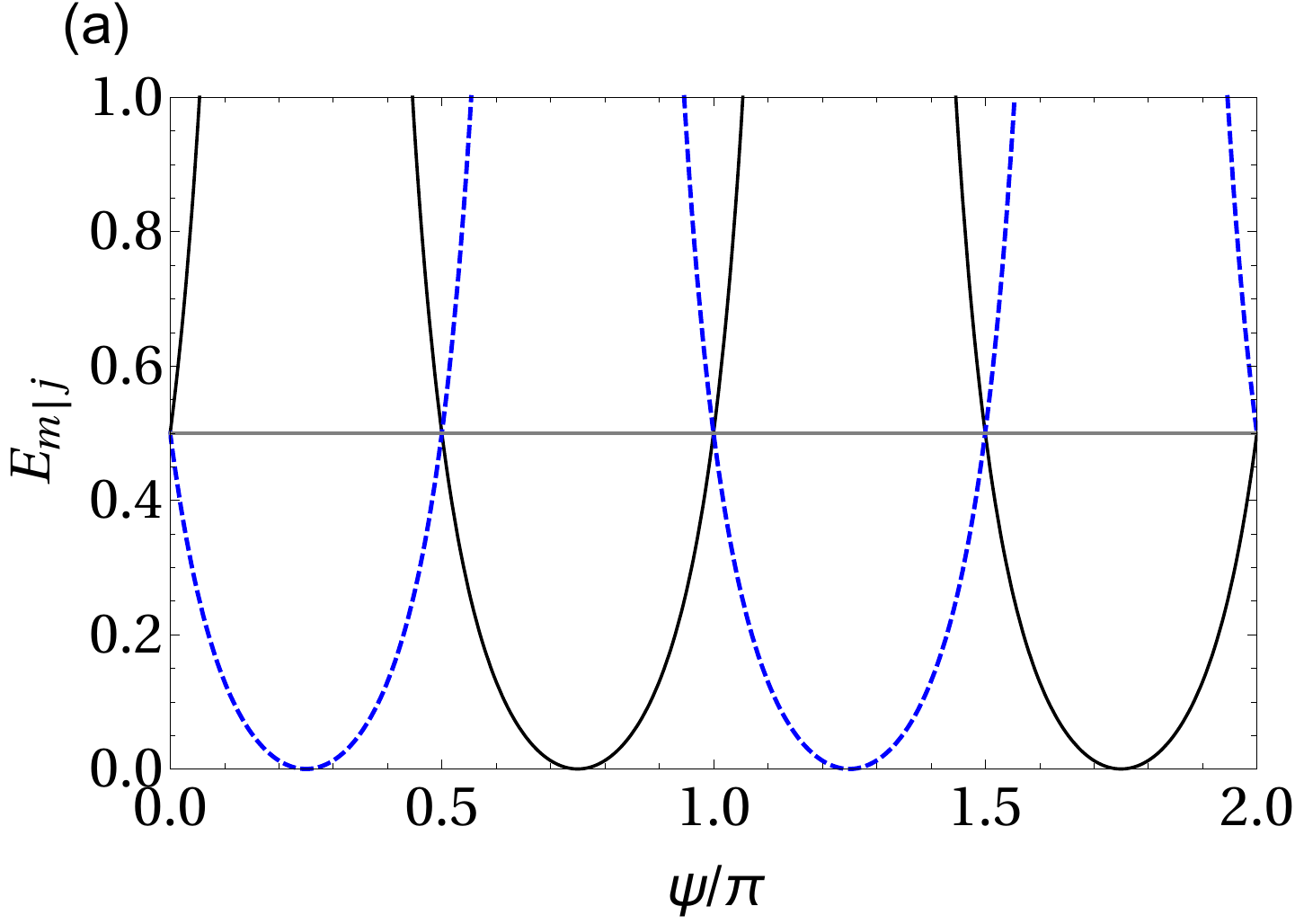}
\includegraphics[width=0.46\textwidth]{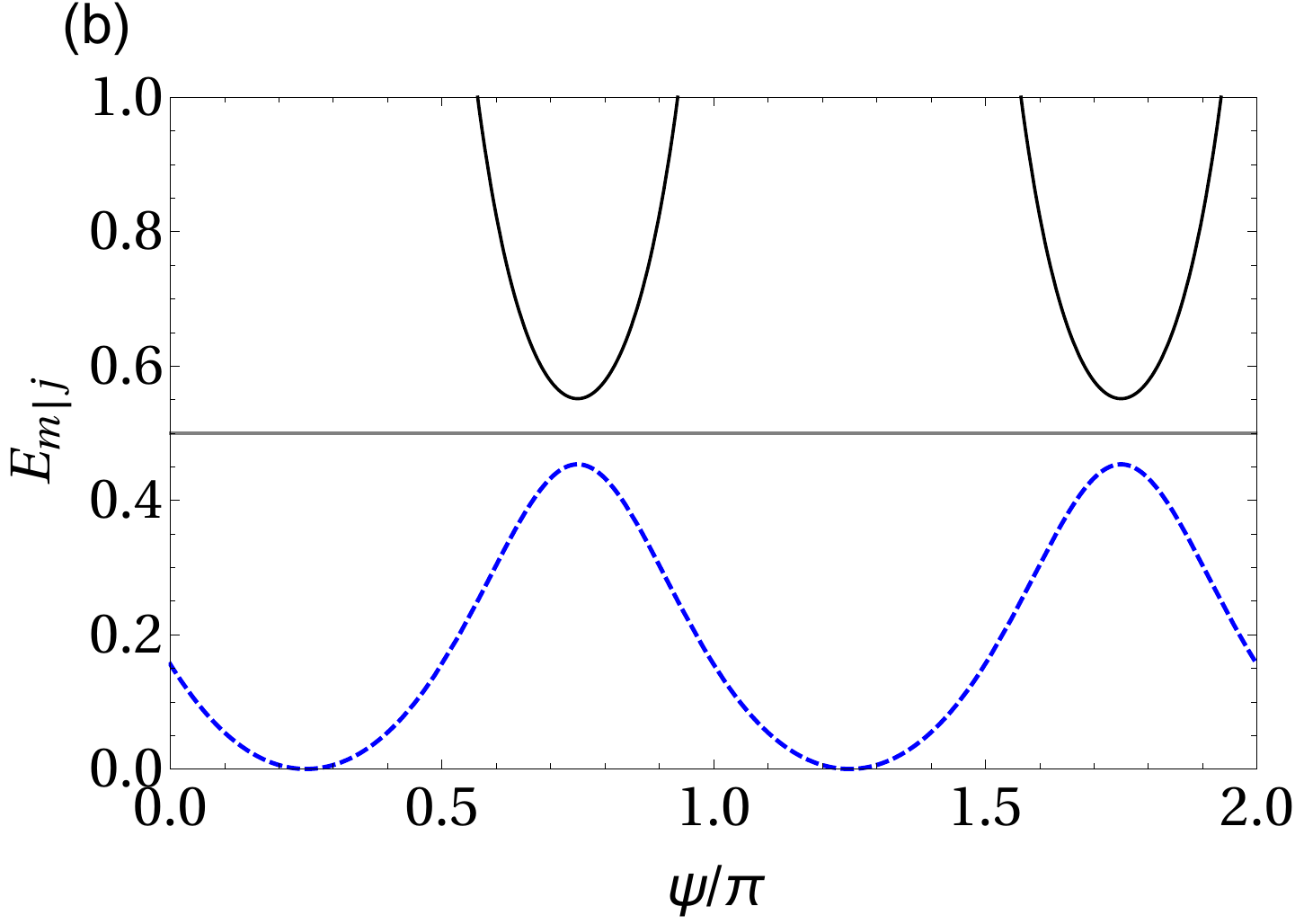}
\caption{Steering parameters $E_{m|1}$ (black solid line) and $E_{m|2}$ (blue dashed line) plotted as a function of the phase difference $\psi$ for $\Delta=0$, $\gamma=0$, $n_{0} =200$. In frame (a) $|g_1|=|g_2|=g$, $J=\kappa=10g$, and in frame (b) $|g_1|\equiv g$, $|g_{2}|= 2.5g$, $\kappa=10g$, $J=4g$. Here, $g=0.1$ MHz.}
\label{fig:6}
\end{figure}

	The above considerations are illustrated in figure~\ref{fig:6}, which shows the variation of the steering parameters $E_{m|1}$ and $E_{m|2}$ with the phase difference $\psi$ for $|g_{1}|=|g_{2}|$ and $|g_{1}|\neq |g_{2}|$. It is seen that for $r\gg 1$ and the phases $\psi$ at which $|{\cal U}_{j}(\psi)|^{2}=1$, the steering parameter $E_{m|j}$ approaches zero that optimum steering is achieved. For the asymmetric case of $|g_{1}|\neq |g_{2}|$ the optimum steering occurs only between the mods $A_{m}^{{\rm out}}$ and $A_{2}^{{\rm out}}$. Thus, we can achieve perfect steering of the mechanical mode by one of the field modes simply through varying the phase difference $\psi$. We can also see that for a given phase $\psi$ only one of $E_{m|1}$ and $E_{m|2}$ can be smaller than $1/2$. In other words, when mode $1$ can steer the mechanical mode, mode $2$ cannot. There is a simple physical interpretation of this feature by referring to the monogamy condition that two or more modes cannot simultaneously steer another mode~\cite{reid2013monogamy}. It is interesting to note that no such rigorous constrain exists for entanglement. As one can see from figures~\ref{fig:3} and \ref{fig:4}, two field modes can be simultaneously entangled with the mechanical mode.

Another interesting feature of the quantum steering present in this system concerns the steering of the mechanical mode by the superposition modes appearing as collective modes. In accordance with the monogamy condition, the presence of the collective steering of the mechanical mode, either $E_{m|u}<1/2$ or $E_{m|w}<1/2$ is accompanied by the absence of the bipartite steering, both $E_{m|1}>1/2$ and $E_{m|2}>1/2$. In other words, collective steering is present in regimes that do not show any bipartite steering. Therefore, the phase dependence of the steering parameters makes the collective steering clearly distinguishable from the bipartite steering. In effect, we can observe perfect one-sided devices independent quantum secret sharing, which is achieved by creating collective steering without creating the bipartite steering~\cite{wang2015efficient}.

Finally, we consider the effect of the damping rate $\gamma$ and the thermal noise $n$ at the mechanical mode on the steering properties. 
Figure~\ref{fig:7} shows the effects of the rate $\gamma$ and the thermal photons $n$ on the optimum value of the steering parameter by taking $|{\cal U}_{j}(\psi)|^{2}=1$. 
The damping $\gamma$ and the thermal noise $n$ affect the optimum steering in a similar way. Interestingly, however, a weak sensitivity to $\gamma$ and $n$ occurs in a region of the squeezing parameter $r$ significantly different from that which maximizes steering in the absence of the damping. Namely, for short pulses $(r<1)$, the effect of $\gamma$ and $n$ is seen to be negligibly small, that $E_{m,j}$ remains quite close to zero over a large range of $\gamma$ and $n$. For $r\gg 1$, the minimum of $E_{m|j}$ degrades more rapidly with $\gamma$ and $n$.

\begin{figure}[h]
\centering
\includegraphics[width=0.46\textwidth]{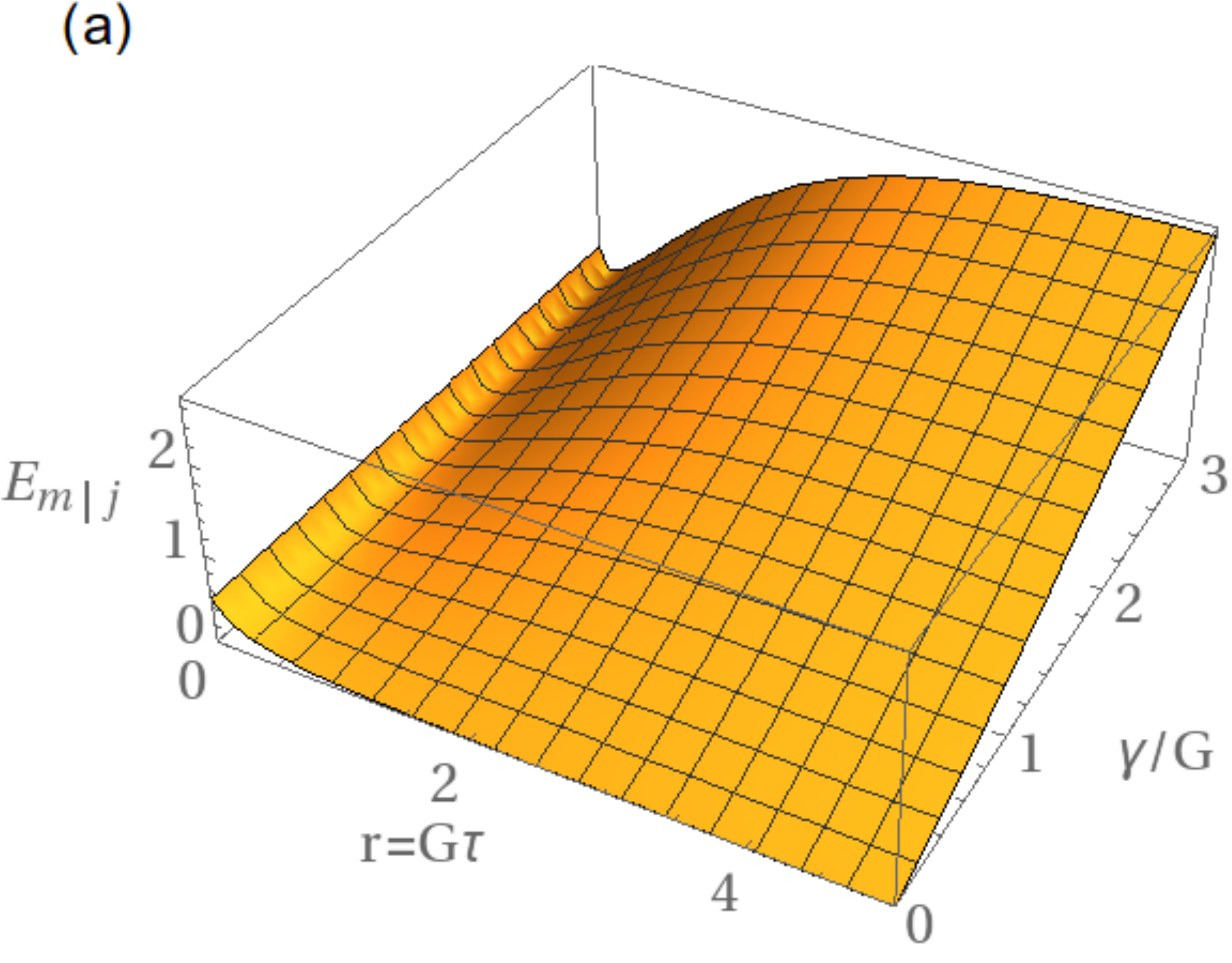}
\hspace{0.02\textwidth}
\includegraphics[width=0.46\textwidth]{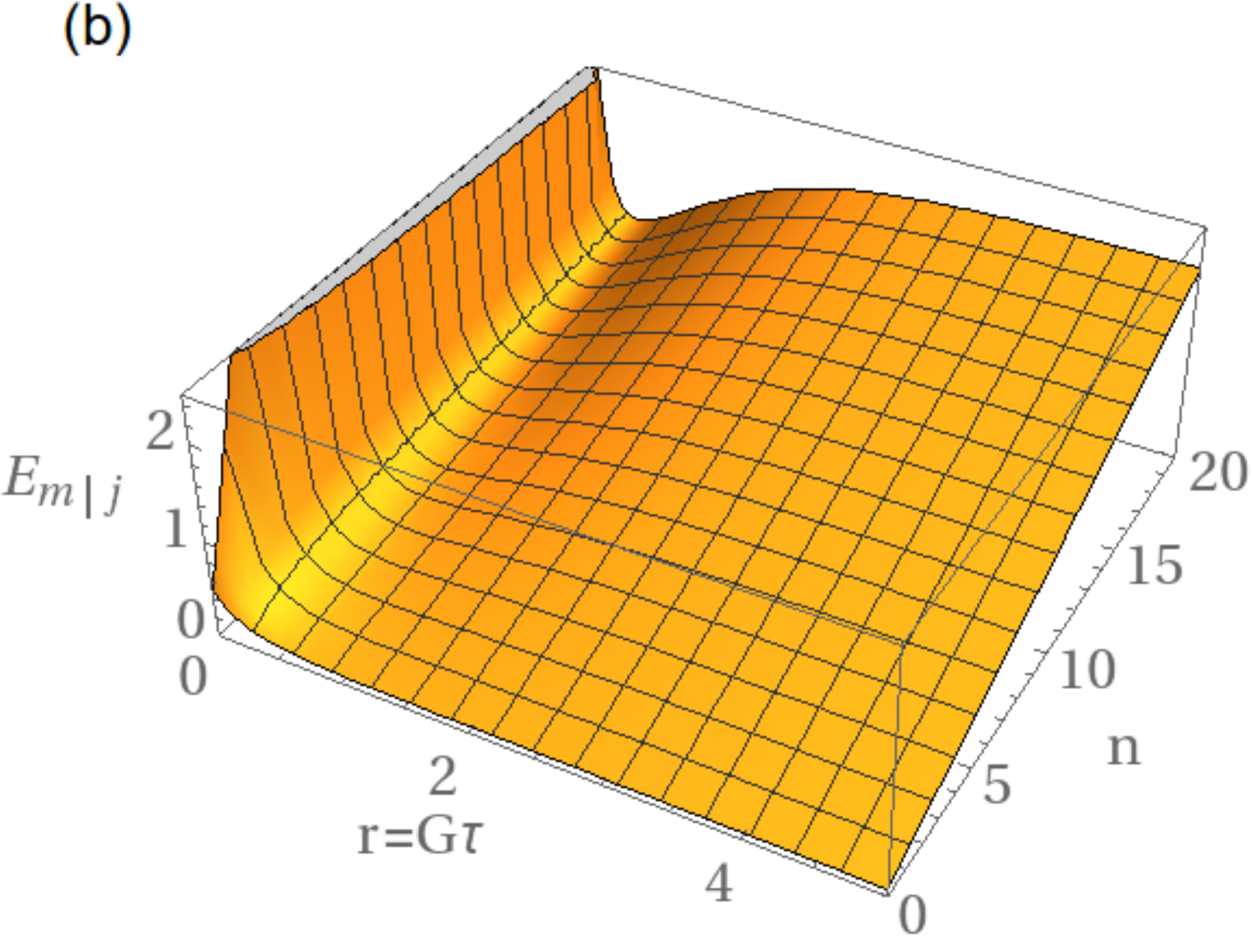}
\caption{Variation of the minimum of the steering parameter $E_{m|j}$ determined by $|{\cal U}_{j}(\psi)|^{2}=1$ in (a) with $r=G\tau$ and $\gamma/G$ for $n_{0}=n_{m}=0$, and in (b) with $r=G\tau$ and $n\equiv n_{0}=n_{m}$ for $\gamma =0.1G$.}
\label{fig:7}
\end{figure}

In concluding this section, we comment about how to test for the presence of bipartite steering of the mechanical mode generated by the cavity modes $1$ and $2$, and the collective steering generated by the modes $u$ and $w$. The generation of the bipartite and collective steerings is ultimately connected to the redistribution of the population among the cavity modes. 
The bipartite steering $E_{m|1}<1/2$ or $E_{m|2}<1/2$ is generated by the population transfer from mode $2$ to $1$ or mode $1$ to $2$ induced by phase difference, respectively. The collective steering, on the other hand, is generated by the equal distribution of the population between the modes $1$ and $2$. One can confirm bipartite and tripartite steerings simply by measuring the coincidence rate $R_{12}$ between 
two detectors $D_{1}$ and $D_{2}$ adjusted to collect photons of the output modes $A_1^{\rm out}$ and $A_2^{\rm out}$, respectively. According to equation~(\ref{III19}), the coincidence rate exhibits a cosine modulation with the phase $\psi$, with the depth of modulation varying between $0$ when one of the output modes, either $A_1^{\rm out}$ or $A_2^{\rm out}$ is not populated, and $2[\kappa g^{2}\Upsilon(r)/G(\kappa^{2} +J^{2})]^{2}$ when they are equally populated. Thus, the minima of the interference pattern of $R_{12}$ signal the bipartite steering, either $E_{m|1}<1/2$ or $E_{m|2}<1/2$, and the maxima signal the collective steering, either $E_{m|u}<1/2$ or $E_{m|w}<1/2$.

\section{Conclusions}\label{sec.6}

We have developed a theory of the phase control of mutual coherence, entanglement, and quantum steering in a system composed of a single mode cavity containing a dielectric membrane in its interior. The membrane was treated as a partly transmitting and reflecting mirror which divides the cavity into two mutually coupled optomechanical cavities. The resulting system has been found to behave effectively as a three-mode linked coupling or closed loop system, the two cavity (optical) modes and the mechanical mode representing the oscillating membrane. The effect of this closed loop is to create interfering channels which depend on the relative phase of the coupling strengths of the field modes to the mechanical mode. The influence of this phase on the population distribution and correlations between the modes has been calculated for the case of short laser pulse excitations of the cavity modes. Manipulation of the phases, which can be done by varying the laser phases, can give rise to a transfer of the population between the cavity modes and to create superposition (collective) modes.  

Our calculations demonstrate that the phase controlled transfer of the population may result in an interesting phenomenon of the induced coherence without induced emission. This phenomenon allows the cavity modes to remain perfectly coherent even if one of the modes is not populated. It has been found that the perfect coherence among the field modes excludes the possibility of the modes to be entangled. On the other hand, it makes possible of the modes to be maximally entangled with the mechanical mode and to behave deterministically that the entanglement can be periodically transferred among the modes by varying the relative phase. 
Furthermore, we have investigated the effect of the relative phase on quantum steering of the mechanical mode by the field modes. It has been found that for a given phase difference only one of the four steering parameters can be smaller than the vacuum level limit. In other words, the mechanical mode can be steered by one of the field modes only. Moreover, we have demonstrated that the phase dependence makes the collective steering clearly distinguishable from the bipartite steering. Such effects can be observed by measuring the coincidence rate between two detectors adjusted to collect photons of the output cavity modes.

The phase dependent dynamics of this close loop system suggest a measurement technique to distinguish between the bipartite and collective steerings. These different types of quantum steering can be observed by measuring the coincidence rate between two detectors adjusted to collect photons of the output cavity modes. In particular, we have found that the minima of the interference pattern of the coincidence rate signal the bipartite steering, while the maxima signal the collective steering. This makes the system suitable to observe perfect one-sided devices independent quantum secret sharing. Finally, we have considered the effect of the damping rate of the mechanical mode on the optimal entanglement and quantum steering and have found that for short pulses the effect of the damping is negligibly small, that the inseparability parameter and the degree of quantum steering remain almost unchanged over a large range of the damping rate. In addition, we have found that the effect of the thermal noise at the mechanical mode on the entanglement and quantum steering is similar to that of the damping of the mode.

\section*{Acknowledgement}

We would like to thank P. Rabl, Z.Q. Yin and Y. Li for helpful discussion, and acknowledge support from National Natural Science Foundation of China (Grants No. 11622428, No. 11274025, No. 61475006, and No. 61675007), Ministry of Science and Technology of China (Grants No. 2016YFA0301302), Q.Y. He thanks the Cheung Kong Scholars Programme (Youth) of China. 

\appendix 
\section{Excitation with a noisy laser}\label{sec.A1}
In this Appendix we discuss the influence of the exciting laser phase fluctuations on the separability parameters. In the discussion of the phase control of mutual coherence, entanglement, and quantum steering we have assumed the exciting laser to be monochromatic. Real lasers are known to posses a finite bandwidth due to the phase and amplitude fluctuations.  Recent studies of the dynamics of optomechanical systems have demonstrated that the finite bandwidth of the exciting laser field due to phase and amplitude fluctuations can considerably affect the results for cavity cooling~\cite{diosi2008laser,rabl2009phase} and entanglement~\cite{abdi2011effect,ghobadi2011optomechanical,dan2013effect}.
Our interest is in the effect of the laser noise on the entanglement. 
When we extend our previous description of the driving laser fields to the case of fluctuating phase, we find that the Hamiltonian of the system takes the form
\begin{eqnarray}
H &=& \hbar\omega_{1}a^{\dag}_{1}a_{1}+\hbar\omega_{2}a^{\dag}_{2}a_{2}+\hbar\omega_{m}c^{\dag}c+\hbar J\left(a_{1}^{\dag}a_{2}+ a_{2}^{\dag}a_{1}\right) \nonumber \\
&+&\hbar (g_{0,1}a^{\dag}_{1}a_{1} + g_{0,2}a^{\dag}_{2}a_{2})(c^{\dag}+c) \nonumber\\
&+&i\hbar\left(E_{01}a_{1}^{\dag}e^{-i\left[ \omega_{L}t+\varphi_{01}+\varphi_{1}(t) \right]} +E_{02}a_{2}^{\dag}e^{-i\left[ \omega_{L}t+\varphi_{02}+\varphi_{2}(t) \right]} - {\rm H.c.}\right) ,\label{H_noise}
\end{eqnarray}
where $\varphi_{01}, \, \varphi_{02}$ are the initial constant parts and $\varphi_{1}(t)$ and $\varphi_{2}(t)$ are the fluctuating parts of the phases of the driving laser fields. Please note that those symbols already defined in the main text will not be redefined in this Appendix.

After going to a rotating frame with $a_{1}\rightarrow a_{1}^{r}\exp\{-i\left[ \omega_{L}t+\varphi_{01}+\varphi_{1}(t) \right]\}$ and $a_{2}\rightarrow a_{2}^{r}\exp\{-i\left[ \omega_{L}t+\varphi_{02}+\varphi_{2}(t) \right]\}$, the Langevin equations are obtained of the form
\begin{eqnarray} 
	\dot{a}_{1}^{r}&=&-(\kappa_{1}+i\Delta_{1}-i\dot{\varphi}_{1})a_{1}^{r}-ig_{0,1}a_{1}^{r}(c+c^{\dagger})-iJa_{2}^{r}e^{i(\delta\varphi+\delta\varphi_{0})}+E_{01}-\sqrt{2\kappa_{1}}a_{1}^{\rm in}, \nonumber\\
	\dot{a}_{2}^{r}&=&-(\kappa_{2}+i\Delta_{2}-i\dot{\varphi}_{2})a_{2}^{r}-ig_{0,2}a_{2}^{r}(c+c^{\dagger})-iJa_{1}^{r}e^{-i(\delta\varphi+\delta\varphi_{0})}+E_{02}-\sqrt{2\kappa_{2}}a_{2}^{\rm in}, \nonumber\\
	\dot{c}&=&-(\gamma+i\omega_{m})c-ig_{0,1}a_{1}^{r\dagger}a_{1}^{r}-ig_{0,2}a_{2}^{r\dagger}a_{2}^{r}-\sqrt{2\gamma}c^{\rm in} ,
\end{eqnarray}
where, for clarity of the notation, we have omitted the time argument of the phases $\varphi_{1}(t)$, $\varphi_{2}(t)$, and $\delta\varphi=\varphi_1(t)-\varphi_2(t)$. Following the same procedure as in references~\cite{abdi2011effect,ghobadi2011optomechanical,dan2013effect}, we apply the linearization and make the rotating-wave approximation to get
\begin{eqnarray}
	\dot{a}_{1}&=&-(\kappa_{1}+i\Delta_{1})a_{1}-ig_{1}c^{\dagger}-iJe^{i\delta\psi}a_{2}+i\dot{\varphi}_{1}\alpha_{1}-\sqrt{2\kappa_{1}}a_{1}^{\rm in}, \nonumber\\
	\dot{a}_{2}&=&-(\kappa_{2}+i\Delta_{2})a_{2}-ig_{2}c^{\dagger}-iJe^{-i\delta\varphi}a_{1}+i\dot{\varphi}_{2}\alpha_{2}-\sqrt{2\kappa_{2}}a_{2}^{\rm in}, \nonumber\\
	\dot{c}&=&-\gamma c-ig_{1}a_{1}^{\dagger}-ig_{2}a_{2}^{\dagger}-\sqrt{2\gamma}c^{\rm in} .\label{A3}
\end{eqnarray}
Here we have used the same rotating farm for equation~(\ref{II6}), $\delta c^m = \delta c e^{i\omega_mt},\ \delta c^{\rm in,m} = \delta c^{\rm in}e^{i\omega_mt}, \ \delta a^{r,m}_{j} = \delta a_{j}^{r}e^{-i(\omega_mt +\varphi_{0j})}$, $\delta a^{\rm in,r,m}_{j} = \delta a^{\rm in,r}_{j}e^{-i(\omega_mt +\varphi_{0j})}$. And for clarity of
the notation, we have omitted the symbol $\delta$ and the superscripts $r$ and $m$ on the displacement operators, as done for equation~(\ref{II6}).

As done in the previous calculations in section~\ref{sec.2}, we assume $\kappa_{1}=\kappa_{2}\equiv\kappa$, $\Delta_{1}=-\Delta_{2}\equiv\Delta$, and taking $\varphi_{1}=\varphi_{2}\equiv\varphi$, $\dot{\varphi}\equiv\sigma$, we then find that in terms of the operators of the superposition modes, equations (\ref{A3}) can be written as
\begin{eqnarray}
	\dot{a}_{w}&=&-\kappa a_{w}-iw a_{w} -ig_{w}c^{\dagger}+i\alpha_{w}\sigma-\sqrt{2\kappa}a_{w}^{\rm in} ,\nonumber\\
	\dot{a}_{u}&=&-\kappa a_{u}+iw a_{u} -ig_{u}c^{\dagger}+i\alpha_{u}\sigma-\sqrt{2\kappa}a_{u}^{\rm in} ,\nonumber\\
	\dot{c}^{\dagger}&=&-\gamma c^{\dagger}+ig_{w}^{\ast}a_{w}+ig_{u}^{\ast}a_{u}-\sqrt{2\gamma}c^{in\dagger} ,
\end{eqnarray}
where $\alpha_{w}=\cos\theta\alpha_{1}+\sin\theta\alpha_{2}$, $\alpha_{u}=\sin\theta\alpha_{1}-\cos\theta\alpha_{2}$.

In the bad cavity limit $\kappa\gg |g_{w,u}|$, we set $\dot{a}_{w}=\dot{a}_{u}=0$, and find
\begin{eqnarray}
	a_{w}&=&-\frac{e^{-i\phi}}{\sqrt{\kappa^{2}+w^{2}}}\left[ ig_{w}c^{\dagger}+\sqrt{2\kappa}a_{w}^{\rm in}-i\alpha_{w}\sigma \right],\nonumber\\
	a_{u}&=&-\frac{e^{i\phi}}{\sqrt{\kappa^{2}+w^{2}}}\left[ ig_{u}c^{\dagger}+\sqrt{2\kappa}a_{u}^{\rm in}-i\alpha_{u}\sigma \right],\nonumber\\
	\dot{c}&=&(G+i\delta)c+i\sqrt{2G_{w}}e^{i\phi}a_{w}^{\rm in\dagger}+i\sqrt{2G_{u}}e^{-i\phi}a_{u}^{\rm in\dagger}-\sqrt{2\gamma}c^{\rm in}\nonumber\\
	&&-\frac{1}{\sqrt{\kappa^{2}+w^{2}}}\left( g_{w}\alpha_{w}^{\ast}e^{i\phi}+g_{u}\alpha_{u}^{\ast}e^{-i\phi} \right)\sigma.
\end{eqnarray}

Incorporating the input-output relation $a^{{\rm out}}=a^{{\rm in}}+\sqrt{2\kappa}a$, we can get the following expressions for the populations and the correlation functions of the normalized temporal field modes
\begin{eqnarray}
	\langle (A_{m}^{{\rm out}})^\dagger A_{m}^{{\rm out}} \rangle &=& n_{0}+\Gamma(r)+|\beta_{wu}|^{2}e^{2r}\langle D^{\dagger}(t)D(t')\rangle ,\nonumber\\
	\langle (A_{w}^{{\rm out}})^{\dagger}A_{w}^{{\rm out}} \rangle &=& \frac{G_{w}}{G}\left\{\Upsilon(r)+\frac{|\beta_{wu}|^{2}e^{2r}}{1-e^{-2r}}\langle D^{\dagger}(t)D(t') \rangle\right. \nonumber\\
	&+&\left. \left| \frac{2\alpha_{w}G}{g_{w}}-\beta_{wu} \right|^{2}\frac{1}{e^{2r}-1}\langle \tilde{D}^{\dagger}(t)\tilde{D}(t') \rangle\right. \nonumber\\
	&+&\left. \frac{e^{2r}}{e^{2r}-1}2{\rm Re}\left[ \beta^{\ast}_{wu}\left( \frac{2\alpha_{w}G}{g_{w}}-\beta_{wu} \right) \langle \tilde{D}^{\dagger}(t)D(t') \rangle \right]\right\} , \nonumber
\end{eqnarray}
\begin{eqnarray}
	\langle (A_{u}^{{\rm out}})^{\dagger}A_{u}^{{\rm out}} \rangle&=&\frac{G_{u}}{G}\left\{\Upsilon(r)+\frac{|\beta_{wu}|^{2}e^{2r}}{1-e^{-2r}}\langle D^{\dagger}(t)D(t') \rangle\right. \nonumber\\
	&+&\left. \left| \frac{2\alpha_{u}G}{g_{u}}-\beta_{wu} \right|^{2}\frac{1}{e^{2r}-1}\langle \tilde{D}^{\dagger}(t)\tilde{D}(t') \rangle\right. \nonumber\\
	&+&\left. \frac{e^{2r}}{e^{2r}-1}2{\rm Re}\left[ \beta^{\ast}_{wu}\left( \frac{2\alpha_{u}G}{g_{u}}-\beta_{wu} \right) \langle \tilde{D}^{\dagger}(t)D(t') \rangle \right]\right\} , \nonumber
\end{eqnarray}
\begin{eqnarray}
	\langle A_{m}^{{\rm out}}A_{w}^{{\rm out}} \rangle e^{-i\phi_{g_{w}}} &=& -i\sqrt{\frac{G_{w}}{G}}\left\{\Lambda(r)+e^{-i\phi}\frac{|\beta_{wu}|^{2}e^{3r}}{\sqrt{e^{2r}-1}}\langle D(t)D^{\dagger}(t') \rangle\right. \nonumber\\
	&&\left.+\, e^{-i\phi}\left( \frac{2\alpha_{w}G}{g_{w}}-\beta_{wu} \right)\frac{\beta^{\ast}_{wu}e^{r}}{\sqrt{e^{2r}-1}}\langle D(t)\tilde{D}^{\dagger}(t')\rangle\right\} , \nonumber
\end{eqnarray}
\begin{eqnarray}
	\langle A_{m}^{{\rm out}}A_{u}^{{\rm out}} \rangle e^{-i\phi_{g_{u}}}&=&-i\sqrt{\frac{G_{u}}{G}}\left\{\Lambda(r)+e^{-i\phi}\frac{|\beta_{wu}|^{2}e^{3r}}{\sqrt{e^{2r}-1}}\langle D(t)D^{\dagger}(t') \rangle\right. \nonumber\\
	&&\left. +\, e^{-i\phi}\!\left( \frac{2\alpha_{u}G}{g_{u}}-\beta_{wu}\!\right)\!\frac{\beta^{\ast}_{wu}e^{r}}{\sqrt{e^{2r}-1}}\langle D(t)\tilde{D}^{\dagger}(t')\rangle\right\} , 
	\label{A6}
\end{eqnarray}
in which
\begin{eqnarray}
\beta_{wu}=\frac{g_{w}^{\ast}\alpha_{w}e^{-i\phi}+g_{u}^{\ast}\alpha_{u}e^{i\phi}}{\sqrt{\kappa^{2}+w^{2}}}, \, D=\int_{0}^{\tau}dt\, \sigma(t)e^{-(G+i\delta)t} ,\, \tilde{D}=\int_{0}^{\tau}dt\, \sigma(t)e^{(G-i\delta)t}.\nonumber \\
\end{eqnarray}
We see that the contribution of the terms resulting from the phase fluctuations are proportional to $|\beta_{wu}|$ and to the noise correlation functions $\langle D^{\dagger}(t)D(t')\rangle$ etc. Thus, the contribution of these terms could be negligibly small if either $|\beta_{wu}|\ll 1$ or $\langle D^{\dagger}(t)D(t')\rangle\ll 1$.
Since $|\beta_{wu}|\sim |\alpha_{i}|,\, (i=1,2)$ and $|\alpha_{i}|\gg 1$, the condition of $|\beta_{wu}|\ll 1$ is not met in our optomechanical system. Therefore, to determine the correlation functions $\langle D^{\dagger}(t)D(t')\rangle$ etc., appearing in equation~(\ref{A6}), we consider a colour-noise model of the excitation laser for which the 
correlation function $\langle \sigma(t)\sigma(t') \rangle$ is given by~\cite{rabl2009phase}
\begin{equation}
	\langle \sigma(t)\sigma(t') \rangle=\Gamma_{l}\gamma_{c}e^{-\gamma_{c}|t-t'|} ,
\end{equation}
where $\Gamma_{l}$ is the laser bandwidth, and the parameter $\gamma_{c}^{-1}$ characterizes a finite correlation time of the phase noise process. The limit $\gamma_{c}\rightarrow \infty$ corresponds to the case of a white-noise laser. 

Note that in the white-noise limit of $\gamma_{c}\rightarrow \infty$, the phase fluctuation terms in equation~(\ref{A6}) become proportional to $|\beta_{wu}|\Gamma_{l}$, which for a finite linewidth $\Gamma_{l}$ could be negligibly small only if $|\beta_{wu}|\ll 1$. As we have already mentioned the condition $|\beta_{wu}|\ll 1$ is not met in the system, so one can expect a significant contribution of these terms to the populations and the correlation functions of the normalized temporal field modes in the white-noise limit.
\begin{figure}[h]
	\centering
	\includegraphics[width=0.46\textwidth]{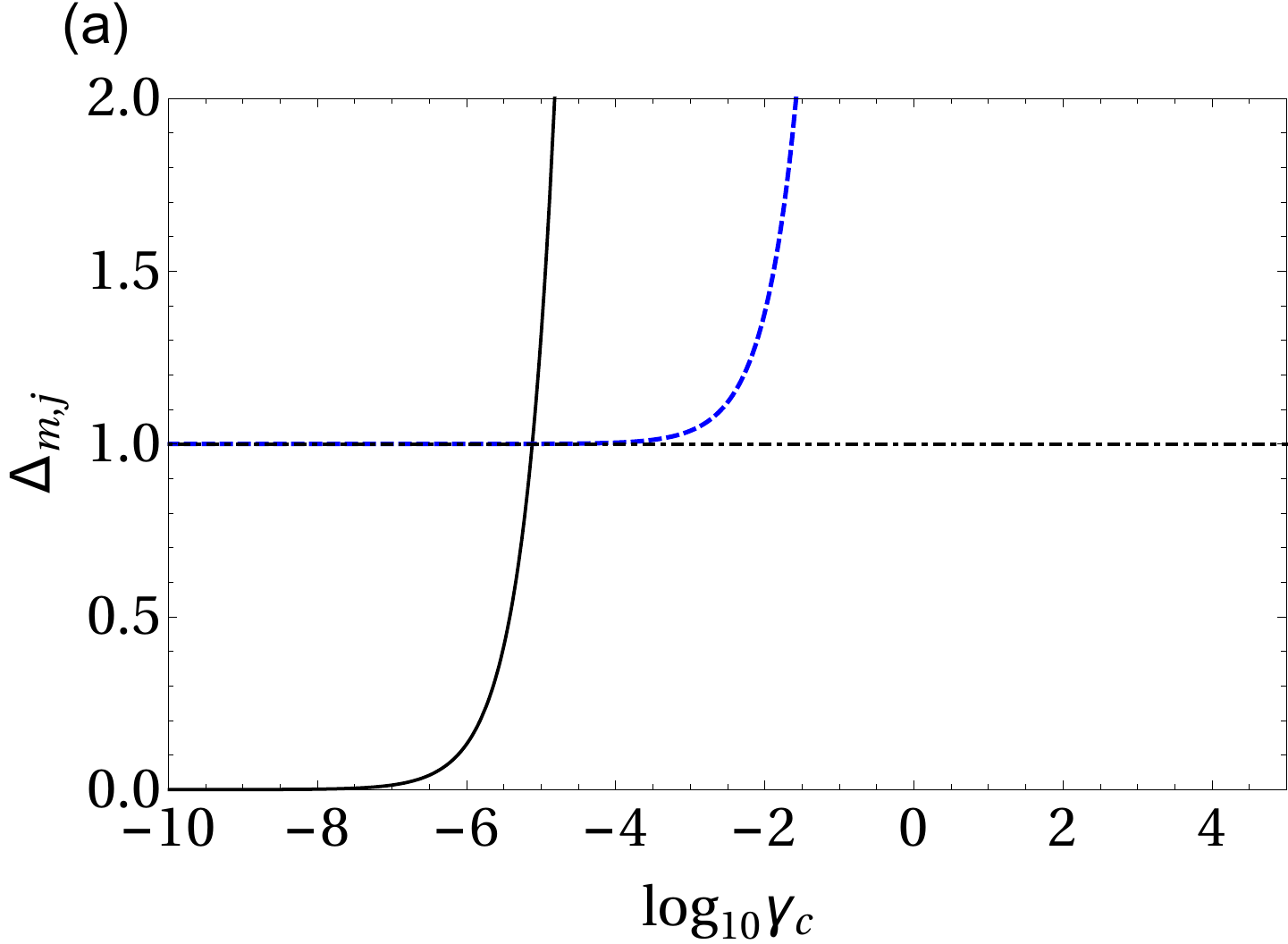}
	\includegraphics[width=0.46\textwidth]{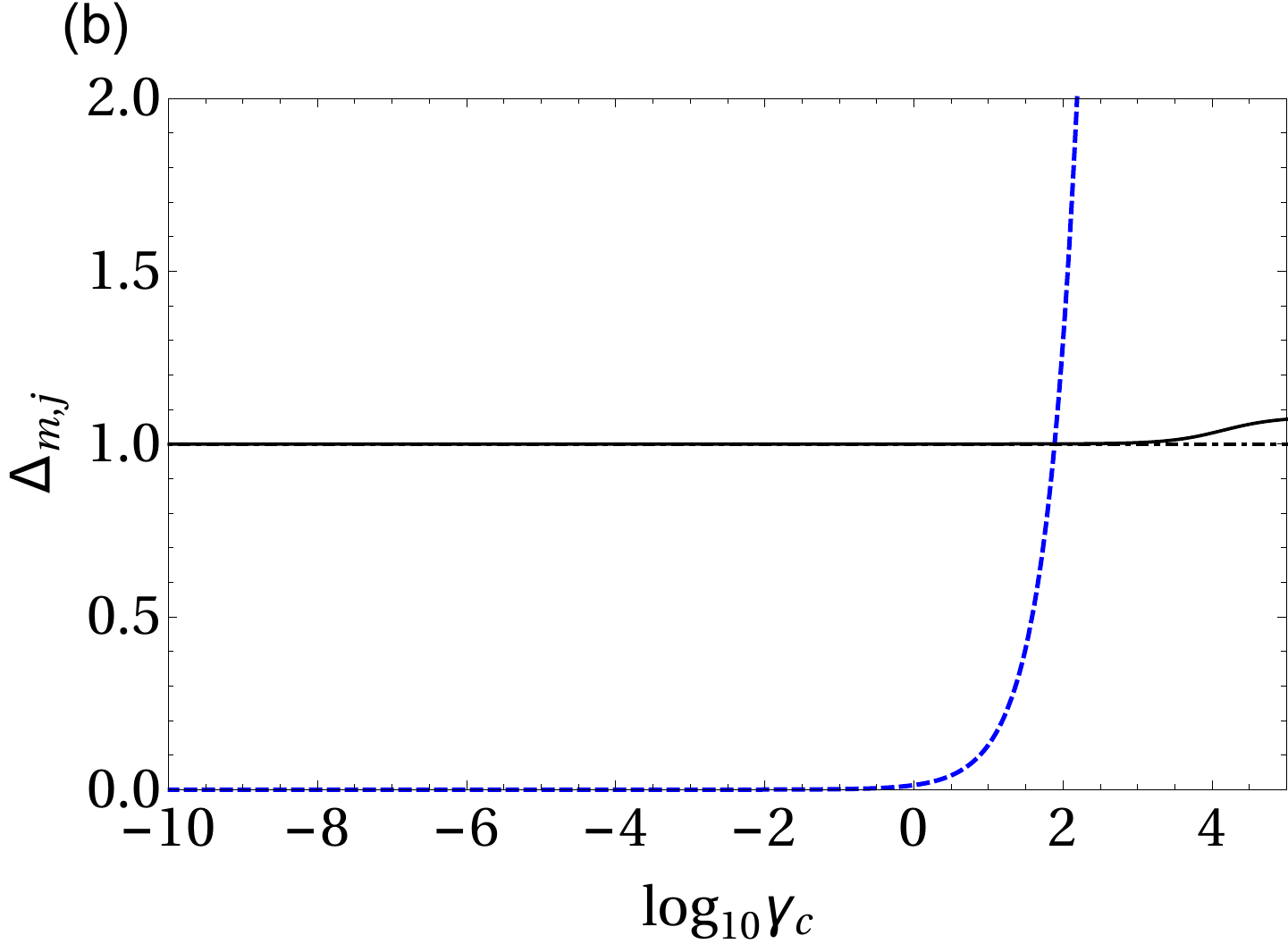}
	\caption{The effect of the laser fluctuations on the minimum values of the inseparability parameters $\Delta_{m,w}$ and $\Delta_{m,u}$ obtained for the case of the monochromatic excitation. The parameters are $g_{0,1}=g_{0,2}=10$ Hz, $|g_1|=|g_2|\equiv g=0.1$ MHz, $J=\kappa=10g$, $\Delta=0$, $\gamma=0$, $n_{0}=n_{m}=200$, $r=G\tau =5$, $\Gamma_{l}=1$ kHz. The black solid line is for $\Delta_{m,w}$, and the blue dashed line is for $\Delta_{m,u}$. In frame (a), $\psi=0$ at which $\Delta_{m,w}$ approaches $0$ in the case of the excitation with a monochromatic laser. In frame (b), $\psi=\pi/2$ at which $\Delta_{m,u}\rightarrow 0$ in the monochromatic case.}
	\label{fig:noise}
\end{figure}

For a colour-noise laser $\gamma_{c}$ could be very small resulting in a negligibly small contribution of the phase fluctuations terms. In order to show this, we evaluate the separability parameters, as given in equation~(\ref{IV8}), but with the correlation functions (\ref{A6}) which include the effect of the fluctuations of the excitation lasers. In particular, we evaluate the effect of the phase fluctuations on the optimal entanglement previously obtained for the monochromatic case and illustrated in figure~\ref{fig:3}(b), i.e., $\Delta_{m,w}=0$ for the phase $\psi$ satisfying $\psi = n\pi, \,(n = 0,1,2,...)$ where the optimal gain $h_w = 1$, and $\Delta_{m,u}=0$ for the phase $\psi$ satisfying $\psi = (n+1/2)\pi, \,(n = 0,1,2,...)$ where the optimal gain $h_u= 1$, respectively.

Figure~\ref{fig:noise} shows the variation of the optimal entanglement with $\gamma_{c}$ as determined by the separability parameters $\Delta_{m,w}$ and $\Delta_{m,u}$ for different relative phases, frame (a) $\psi=0$, and frame (b) $\psi=\pi/2$.  One can see that the optimal entanglement between field mode $A^{{\rm out}}_{w}$ and the mechanical mode $A_{m}^{{\rm out}}$ rapidly disappears at $\gamma_{c}\sim 10^{-5}$ Hz, while the optimal entanglement between the modes $A_{u}^{{\rm out}}$ and $A_{m}^{{\rm out}}$ disappears at $\gamma_{c}\sim 10^{2}$ Hz. The reason why the separability parameters $\Delta_{m,w}<1$ and $\Delta_{m,u}<1$ have different thresholds for $\gamma_{c}$ is that $\alpha_{u}\ll\alpha_{w}$ determined by the considered parameters $|g_{1}|=|g_{2}|$, $\Delta=0$, $\kappa=J$. We may conclude that similar to the laser phase fluctuation effects on the optomechanical cavity cooling~\cite{diosi2008laser,rabl2009phase} and on the entanglement~\cite{abdi2011effect,ghobadi2011optomechanical,dan2013effect}, the optimal entanglement generated in this scheme, corresponding the perfect EPR state, can be experimentally observed if a colour rather than white-noise laser is used to excite the cavity modes.

\section*{References}
\bibliography{Phase_control_correlations_final}
\end{document}